\documentclass[12pt, a4paper]{article}

\usepackage{graphicx}
\usepackage[left = 1.5cm, right = 1.5cm, top = 1.5cm, bottom = 1.5cm, footskip = 0.5cm]{geometry}

\usepackage{amsmath}
\usepackage{url}
\usepackage{siunitx}
\usepackage{multirow}
\usepackage{enumitem}
\usepackage{algorithm}
\usepackage{algpseudocode}
\usepackage{amsfonts}
\usepackage{arydshln}
\usepackage{authblk}
\newtheorem{theorem}{Theorem}
\newtheorem{remark}{Remark}

\usepackage[backend = biber, style = apa]{biblatex}
\usepackage{setspace}
\title{MSR-CP: A Modified Segmented Regression Approach to Change Point Estimation in Rainfall Patterns in West Bengal}
\author[1]{Anirban Ghosh{\thanks{Corresponding author: Anirban Ghosh, Applied Statistics Unit, Indian Statistical Institute, 203 B.T. Road, Kolkata - 700108, India. Email: ganirban004@gmail.com}}}
\author[2]{Raju Maiti}
\affil[1]{Applied Statistics Unit, Indian Statistical Institute, 203 BT Road, Kolkata - 700108, India}
\affil[2]{Economic Research Unit, Indian Statistical Institute, 203 BT Road, Kolkata - 700108, India}

\date{}

\begin{document}

\maketitle

\begin{abstract}
    \noindent
    This study presents a methodology for detecting change point in precipitation series using a smooth version of segmented regression model. We examine several existing change point detection methods and identify their limitations in capturing gradual changes commonly observed in climate data. To address these limitations, we discuss an extension of the traditional segmented regression model enabling the detection of gradual shifts in trend. The developed methodology is further extended to accommodate dependent error structures, which are common in time series data. Extensive simulation studies are conducted to evaluate the performance and computational efficiency of the proposed approach relative to existing methods. The methodology is subsequently applied to 60 years of monsoon rainfall data from two districts of West Bengal, India. This approach offers improved insights into evolving rainfall patterns and serves as a useful tool for climate data analysis and long-term forecasting.
\end{abstract}


\section{Introduction}
A change point problem in time series analysis arises when the underlying data generating mechanism experiences a structural change at an unknown point in time. Such changes may occur in the mean, variance, trend or other characteristics of the time series, reflecting changes in the underlying process. That means the data generating process is expected to change at a given time point, which is generally unknown. Formally, a general formulation of the change point problem for a time series $\{ Y_{t}: t = 1, 2, \ldots, n \}$ is given by
\begin{equation*}
    Y_{t} \sim \left \{ \begin{array}{ll}
		f_{1} & \text{for}\ t \le t_{0}\  \\
		f_{2} & \text{for}\ t > t_{0}\
	\end{array}
	\right.
\end{equation*}
where $f_{1}$ and $f_{2}$ are the underlying probability distribution, generating the data, before and after the change point $t_{0}$. The objective is to estimate the change point $t_{0}$, along with the functional form of two regimes, i.e., $f_{1}(\cdot)$ and $f_{2}(\cdot)$, from the observed data. To find the change in means of the distributions, a common approach is to model the mean, $\operatorname{E}(Y_{t})$, as a piecewise linear regression function before and after the change point $t_{0}$ as
\begin{equation*}
    \operatorname{E}(Y_{t}) = \left \{ \begin{array}{ll}
		\alpha_{1} +\beta_{1} t & \text{for}\ t \le t_{0}\  \\
		\alpha_{2} + \beta_{2} t & \text{for}\ t > t_{0}\
	\end{array}
	\right..
\end{equation*}
But the unknown change point $t_{0}$ enters into the model as a nonlinear parameter making the estimation problem a nonlinear regression problem.

Nonlinear regression models have been widely used in practice for a variety of time series applications (\cite{drignei2008}; \cite{zhang2012}; \cite{chattopadhyay2023}). In the context of change point analysis, the smooth transition regression (STR) models, initially proposed by \cite{chan1986}, are further developed in \cite{luukkonen1988}, \cite{terasvirta1994, terasvirta1996}, \cite{li2002}, \cite{saikkonen2004}, \cite{medeiros2008} and \cite{areosa2011}. \cite{mira2000} discussed the asymptotic properties of nonlinear least squares (NLS) under STR setup with i.i.d. errors. \cite{farinas2004} and \cite{medeiros2005} used smooth transition autoregressive (STAR) models into neural network setup. \cite{hillebrand2013} found the asymptotic properties of quasi-maximum likelihood estimator (QMLE) considering time as a transition variable when the errors are NIID. More recently, \cite{chattopadhyay2022} addressed the issue of change point detection for count time series data by employing an INAR(1) process with time-varying covariates under STR setup. \cite{lu2023} proposed a piecewise segmented regression model, allowing for discontinuous jumps and connected change points in a single model. \cite{yang2026} combined STAR model into interval time series analysis. \cite{joneus2026} combined STR model into hazard regression model. But none of the existing studies studied smooth differentiable version of the segmented regression model under dependent error structure.

The study of climate data, specifically rainfall, is essential for several reasons. Firstly, agricultural prosperity of any region in India heavily depends on timely and adequately distributed rainfall, especially during monsoon season. West Bengal is the highest rice production state of India and commands the lion’s share of India’s rice production,  contributing a substantial 14\% (\cite{rice_production}) to the nation’s overall output. Rice yield, in particular in West Bengal, heavily depends on the timely and adequately distributed monsoon rainfall. Secondly, it helps in predicting and managing water-related disasters like flooding and landslides, which are expected to worsen in India due to climate warming (\cite{nayak2023}). So rainfall patterns directly impact agriculture, water security and the overall economy of agrarian societies like India (\cite{nair2018}). Understanding shifts in rainfall patterns is crucial for improving weather prediction models, optimizing water management strategies and developing effective climate change adaptation measures. In this context, detecting change points in precipitation patterns using various statistical methods has drawn significant attention in recent times. Change point detection in climate data has been carried out using tests such as Buishand Range (BR), Standard Normal Homogeneity (SNH), Pettitt's and the Von Neumann Ratio (VNR), in addition to trend analysis based on the Mann-Kendall (MK) test (\cite{croitoru2012}; \cite{zarenistanak2014}; \cite{jaiswal2015}; \cite{onyutha2016}; \cite{javari2016}; \cite{suhaila2018}). \cite{kundu2019} conducted an analysis of long-term annual and seasonal rainfall trends, along with change point detection of annual rainfall, using 102 years (1901--2002) of monthly rainfall data from 18 stations across West Bengal. Similarly, \cite{getahun2021} employed various parametric and nonparametric approaches for detecting change points in precipitation and temperature data across Ethiopia, analyzing the data on annual, seasonal and monthly scales. \cite{bouregaa2023} further performed change point detection and trend analysis of drought patterns in Algeria from 1901 to 2018.

In this paper, we study an extension of the segmented regression model that incorporates a smooth and differentiable transition function based on \cite{fong2017a}. We investigate the asymptotic properties of the modified version, which is a particular case of STR model, under i.i.d. errors. The derivative-based optimization approach for the modified model requires lower computation time relative to the grid-search based approach for the segmented regression model. We further extend the study to accommodate AR($p$) error structure into the modified smooth model. We propose an estimation procedure based on generalized least square (GLS). We also discuss the asymptotic properties of the estimated parameters. Simulation studies are conducted to demonstrate the applicability of the discussed methodologies in different situations for $p = 1$. To illustrate their practical utility, we apply the proposed change point detection methods to monsoon precipitation data from two districts (South 24 Parganas and Coochbehar) of West Bengal, India.

The remainder of this paper is organized as follows. In Section \ref{sec:data_examples}, we introduce real-world precipitation data of South 24 Parganas and Coochbehar districts of West Bengal. A brief overview of the existing models employed for change point analysis are provided in Section \ref{sec:methods}. In Section \ref{sec:ssrm}, we present a comprehensive discussion of the smooth differentiable version the segmented regression model and its' asymptotic properties. This model is extended for AR($p$) errors in Section \ref{sec:ssrm_arp}. Section \ref{sec:simulation} details the simulation experiments. The change point analysis of the real precipitation data of South 24 Parganas and Coochbehar are presented in Section \ref{sec:data_analysis}. Finally, we offer concluding remarks in Section \ref{sec:conclusion}. All computations and visualizations are conducted using R v4.6.1 (\cite{r2026}). The R codes are available in author's GitHub repository: \url{https://github.com/ganirban004/change-point-paper.git}.

\section{Motivating Precipitation Data}      \label{sec:data_examples}
\begin{figure}[htpb]
    \centering
    \includegraphics{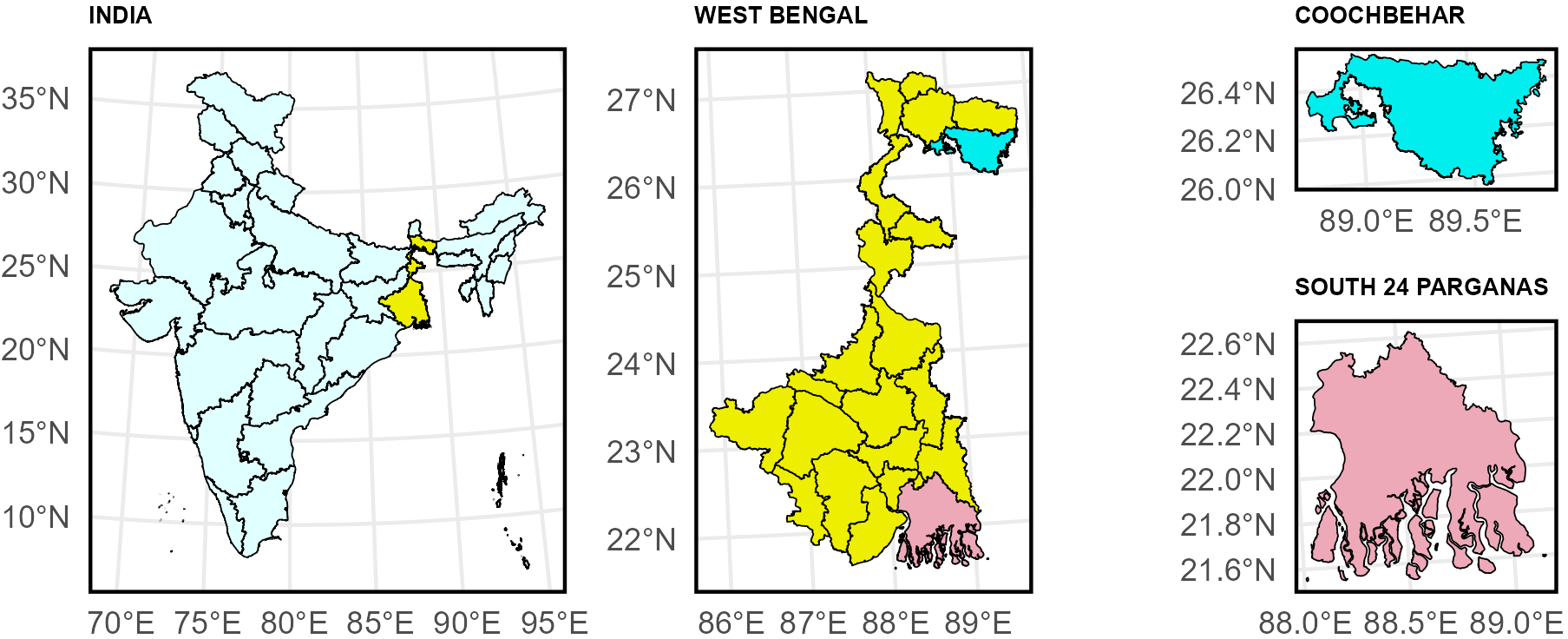}
    \caption{Location map of Coochbehar and South 24 Parganas districts in India}
    \label{fig:map}
\end{figure}

This study analyzes the yearly monsoon precipitation data of South 24 Parganas and Coochbehar districts in West Bengal, India. The Indian state West Bengal is geographically diverse, encompassing the Himalayan foothills in the north the plateau regions in the west, fertile plains in the central region and coastal areas along the Bay of Bengal. The state's climate varies from subtropical in the plains to temperate in the hills, with monsoon-driven rainfall providing the bulk of its water resources, crucial for agriculture and sustenance. Among six agro-climatic zones in West Bengal, South 24 Parganas and Coochbehar belong to the Coastal Saline Zone and Terai-Teesta Alluvial Zone respectively. South 24 Parganas lies between latitudes \ang{21; 53}N to \ang{22; 63}N and longitudes \ang{88; 01}E to \ang{89; 15}E. Coochbehar lies between latitudes \ang{25; 97}N to \ang{26; 55}N and longitudes \ang{88; 75}E to \ang{89; 86}E. The geographical positions of South 24 Parganas and Coochbehar are illustrated in Figure \ref{fig:map}.

\subsection{Dataset Description}
The data are sourced from the West Bengal Pollution Control Board (WBPCB). The dataset consists of observations from multiple rain-gauge stations located across various districts of West Bengal, with each station recording daily precipitation data spanning for 120 years (1901--2020).

To facilitate district-level analysis, we aggregate the station-wise daily data to district-wise daily data by taking the simple average. Suppose there are $m$ stations in a district and $Y_{it}$ denotes the amount of rainfall in station $i$ on day $t$. Then the district-level rainfall for day $t$ is calculated as follows:
\begin{equation*}
    Y_{t} = \frac{1}{m} \sum_{i = 1}^{m} Y_{it}.
\end{equation*}

Similarly, to compute the monthly rainfall, we aggregate the total rainfall over the entire month, yielding the monthly rainfall for the district. The rainfall amounts for June, July, August and September are then aggregated to obtain the annual monsoon rainfall. In this analysis, we focus on the yearly monsoon rainfall for the last 60 years (1961--2020) for South 24 Parganas and Coochbehar districts to study the change point methodology described in this article. The ready-to-use monsoon precipitation data of South 24 Parganas and Coochbehar for last 60 years are available from the corresponding author upon request. We present an exploratory analysis of the data of South 24 Parganas in the next section, while a similar exploratory analysis for Coochbehar data is presented in Appendix \ref{app:eda_coochbehar} for the sake of brevity.

\subsection{Exploratory Data Analysis}  \label{sec:eda}
In this section, we conduct an exploratory data analysis of the monsoon precipitation series for South 24 Parganas to identify underlying patterns, trends and structural features in the data, which subsequently motivates the change point analysis performed in later sections of this article. During the period 1961--2020, the mean monsoon rainfall in South 24 Parganas is found to be 1457.89 mm, with a standard deviation of 357.45 mm, indicating substantial inter-annual variability. The highest precipitation was recorded in 1981, amounting to 2843.35 mm, while the lowest rainfall occurred in 1976, with a total of 783.85 mm.

\begin{figure}[htpb]
    \centering
    \includegraphics{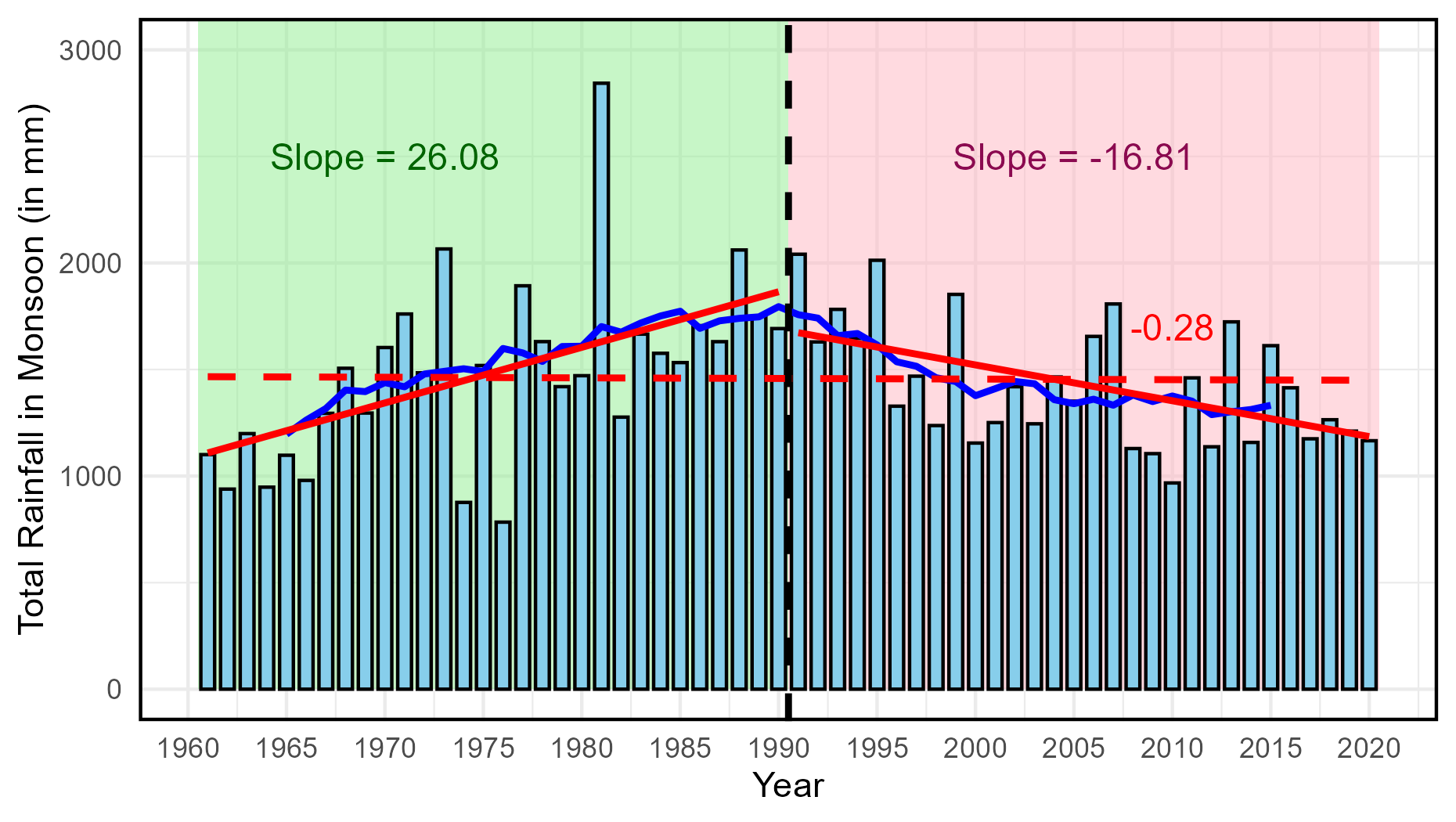}
    \caption{Yearly total precipitation (in mm) in monsoon in South 24 Parganas. The blue curve indicates 10-year rolling means. The red lines indicate linear trends before and after 1990. The red dashed line is the overall linear trend throughout 1961--2020.}
    \label{fig:trend_s24pgs}
\end{figure}

\begin{figure}[htpb]
    \centering
    \includegraphics{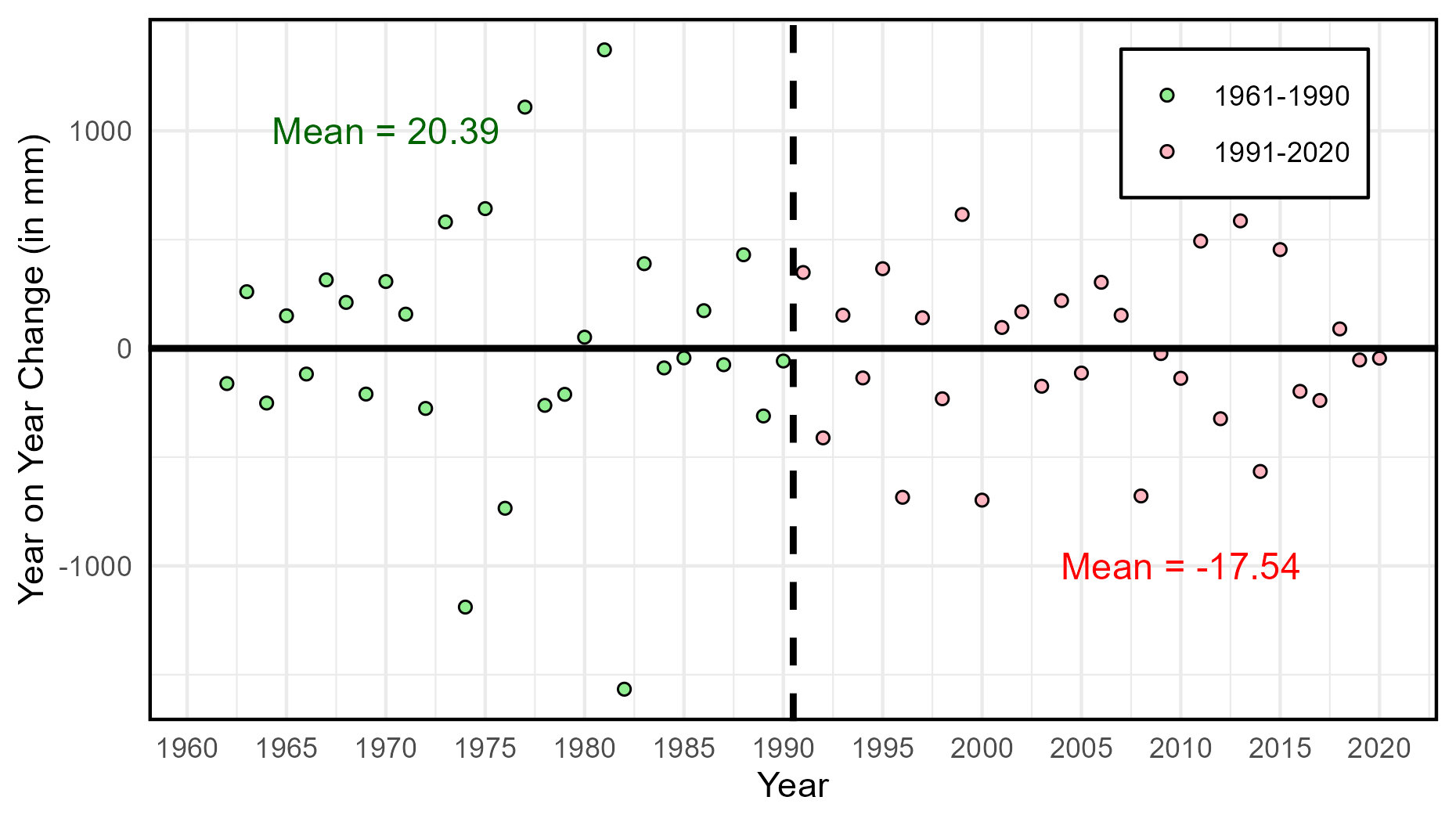}
    \caption{Year-on-year (YoY) change in total precipitation (in mm) in monsoon in South 24 Parganas}
    \label{fig:yoy_s24pgs}
\end{figure}

Figure \ref{fig:trend_s24pgs} depicts the temporal evolution of annual monsoon rainfall in South 24 Parganas. The 10-year rolling mean (blue curve) suggests a declining trend after 1990, whereas it is increasing before 1990. The red lines are linear trend lines fitted separately for the two regimes (upto 1990 and after 1990). It shows a 164.46\% decline in the slope of regression lines after 1990. It suggest a possible trend change in the rainfall regime. Motivated by this observation, we calculate year-on-year (YoY) change as
\begin{equation*}
    \Delta Y_{t} = Y_{t} - Y_{t - 1}
\end{equation*}
where $Y_{t}$ is the total precipitation in monsoon in year $t$ and plot it in Figure \ref{fig:yoy_s24pgs}. It reveals that in the first period (1961--1990), the rainfall in monsoon was increasing at a rate of 20.39 mm per year on an average, whereas in the later period (1961--2020), the monsoon precipitation was decreasing at a rate of 17.54 mm per year on an average. The regime-wise summary statistics are presented in Table \ref{tab:statistics_s24pgs}. Results show a 186\% decline in average YoY change in later period relative to the prior.

The exploratory analysis for Coochbehar data is presented in Appendix \ref{app:eda_coochbehar}. Collectively, the exploratory analyses provide a strong evidence of a change in the monsoon rainfall trend around the year 1990 for both the districts South 24 Parganas and Coochbehar. This motivates a formal investigation of structural changes and consequently, change point estimation is pursued in the subsequent sections of this article.

\begin{table}[htpb]
\centering
\caption{Summary statistics of annual monsoon rainfall (in mm) for two sub-periods, 1961--1990 and 1991--2020, in South 24 Parganas district}
\begin{tabular}{|c|c|c|}
\hline
\textbf{Statistics} & \textbf{1961--1990} & \textbf{1991--2020} \\ \hline
Mean               & 1486.73 & 1429.06 \\ 
Standard Deviation & 417.25 & 282.47 \\ 
Maximum (Year)     & 2843.35 (1981) & 2040.92 (1991) \\ 
Minimum (Year)     & 783.85 (1976) & 967.70 (2010) \\
Average YoY Change & 20.39 & $-$17.54 \\
\hline
\end{tabular}
\label{tab:statistics_s24pgs}
\end{table}

\section{Existing Methods for Detecting Change Point}   \label{sec:methods}
As discussed in Section \ref{sec:eda}, the exploratory analysis suggests the presence of a potential change point in the monsoon rainfall series of both the districts. In this section, we provide a comprehensive overview of widely used single change point detection methodologies, which includes parametric, nonparametric and regression-based approaches, along with their inherent limitations. These established methods form the methodological foundation against which the proposed approach is developed and evaluated.

Suppose $\{Y_{t}\}_{t = 1}^{n}$ be a time series where $Y_{t}$ is the total precipitation in monsoon in year $t$, $t = 1, 2, \ldots, n$. Under this setup, the existing methods are discussed below.

\subsection{Parametric Approaches}
In parametric approaches, each observation $Y_{t}$, $t = 1, 2, \ldots, n$ is assumed to follow a Normal distribution. The considered model with a single shift (change point) is 
\begin{equation*}
    Y_{t} = 
    \begin{cases}
        \mu + \varepsilon_{t}, & t = 1, \ldots, t_{0} \\
        \mu + \Delta + \varepsilon_{t} & t = t_{0} + 1, \ldots, n 
    \end{cases}
\end{equation*}
with $\varepsilon_{t} \sim N(0, \sigma^2)$, $\sigma^{2} > 0$. Then the null hypothesis $H_{0}: \Delta = 0$ is tested against the alternative $H_{1}: \Delta \neq 0$.

\subsubsection{Buishand Range (BR) Test}
Buishand Range test (\cite{buishand1982, verstraeten2006}) is an widely applicable method for testing and detecting a point of mean shift in a time series data. Under the above setup, this method calculates the rescaled adjusted partial sums as
\begin{equation*}
    S_{t} = \sum_{i=1}^t (Y_{i} - \bar{Y})
\end{equation*}
for $t = 1, 2, \ldots, n$, where $\bar{Y}$ is the mean rainfall. The test statistic is calculated as
\begin{equation*}
    R = \frac{\underset{t}{\max} S_{t} - \underset{t}{\min} S_{t}}{\sigma}.
\end{equation*}
Critical values based on 19999 Monte Carlo simulations are tabulated for $\frac{R}{\sqrt{n}}$ in \cite{buishand1982}. The estimated change point is given by
\begin{equation*}
    \hat{t}_{0} = \operatorname*{argmax}_{1 \le t \le n} |S_{t}|
\end{equation*}

The Buishand range test is more sensitive for breaks in the middle of the series (\cite{verstraeten2006}). For abrupt changes in the beginning and at the end of time series, a more appropriate technique is the Standard Normal Homogeneity Test (SNHT).

\subsubsection{Standard Normal Homogeneity (SNH) Test}
Among parametric approaches, Standard Normal Homogeneity test (\cite{alexandersson1986, verstraeten2006}) is one of the most applicable techniques for estimating change point in a time series. Under the above setup, $T_{t}$ is defined as
\begin{equation*}
    T_{t} = tz_{1}^{2} + (n - t)z_{2}^{2}
\end{equation*}
for $t = 1, 2, \ldots, n - 1$, where
\begin{equation*}
    z_{1} = \frac{1}{t}\sum_{i = 1}^{t} \frac{{Y_{i}} - \bar{Y}}{\sigma}, \quad 
z_{2} = \frac{1}{n - t}\sum_{i = t + 1}^{n} \frac{Y_{i} - \bar{Y}}{\sigma}.
\end{equation*}
Then the test statistic to test the null hypothesis $H_{0}: \Delta = 0$ is given by
\begin{equation*}
    T = \underset{t}{\max} T_{t}
\end{equation*}
Critical values based on 1,000,000 Monte Carlo simulations are tabulated for $T$ by \cite{khaliq2007}. The estimated change point is
\begin{equation*}
    \hat{t}_{0} = \operatorname*{argmax}_{1 \le t < n} T_{t}
\end{equation*}

Like most parametric approaches, the BR test and the SNH test rely on the assumptions about the distribution of $Y_{t}$. Violation of distributional assumptions may lead to biased or misleading results. To relax this assumption and achieve greater robustness, nonparametric methods provide suitable alternatives.

\subsection{Nonparametric Approaches}
\subsubsection{Pettitt's Test}
Pettitt’s test (\cite{pettitt1979, verstraeten2006}) is a nonparametric rank-based method used to detect a single point of shift in the location parameter of a time series. The null hypothesis assumes that $Y_{1}, Y_{2}, \ldots, Y_{t_{0}}$ has a distribution function $F_{1}(y)$ similar with $F_{2}(y)$, the distribution function of $Y_{t_{0} + 1}, Y_{t_{0} + 2}, \ldots, Y_{n}$, i.e., $H_{0}: t_{0} = n$. To test the null hypothesis, this method first calculates
\begin{equation*}
    U_{t} = 2 \sum_{i = 1}^{t}r_{i} - t(n + 1)
\end{equation*}
for $t = 1,2 , \ldots, n$, where $r_{i}$ is the rank of the $i^{th}$ observation when the values $Y_{1}, Y_{2}, \ldots, Y_{n}$ in the time series are ranked in ascending order. Then the test statistic is calculated as
\begin{equation*}
    U = \underset{t}{\max} |U_{t}|,
\end{equation*}
and the $p$-value for a two-sided test is calculated as $p = 2 \exp(\frac{-6\hat{t}_{0}^{2}}{n(n + 1)})$, where $\hat{t}_{0}$ is the estimated change point, given by
\begin{equation*}
    \hat{t}_{0} = \operatorname*{argmax}_{1 \le t < n} U_{t}
\end{equation*}

\subsubsection{Sequential Mann-Kendall (SeqMK) Test}
The Sequential Mann--Kendall test (\cite{kendall1990, schaber2005}) is a nonparametric method for detecting trend changes and identifying multiple possible change points without assuming any distribution for the data. For $t = 2, 3, \ldots, n$, the progressive Mann--Kendall statistic is defined as
\begin{equation*}
    S_{t} = \sum_{i = 1}^{n - 1} \sum_{j = i + 1}^{n} \operatorname{sgn}(Y_{j} - Y_{i})
\end{equation*}
where $\operatorname{sgn}(\cdot)$ denotes the sign function. Under the null hypothesis of no trend, $\operatorname{E}(S_{t}) = 0$ and $\operatorname{Var}(S_{t}) = \frac{t(t - 1)(2t + 5)}{18}$, assuming no ties. The progressive sequence is calculated as
\begin{equation*}
    UF_{1} = 0, \ UF_{t} = \frac{S_{t}}{\sqrt{\operatorname{Var}(S_{t})}}, \ t = 2, 3, \ldots, n
\end{equation*}

Applying the same procedure to the reversed series yields the retrograde sequence $\{UB_{t}\}_{t = 1}^{n}$. A potential change point is indicated by the intersection of the $UF_{t}$ and $UB_{t}$ curves. This test detects an intersection point as a change point when either the progressive or the retrograde row exceeds certain confidence limits ($\pm$1.96 for 5\% significance level) before and after the intersection point.

These aforementioned change point detection methods are primarily designed to identify structural breaks characterized by abrupt shifts in the mean. However, such sudden mean changes are not commonly observed in climate variables, including precipitation series for regions such as Coochbehar. Climate data typically exhibit gradual and systematic changes in trend over time, which these methods are not well suited to detect. To effectively capture such gradual trend changes in monsoon rainfall, regression-based change point techniques provide a more appropriate framework. These approaches model structural changes by fitting separate regression relationships on either side of a candidate change point. The regression-based methodologies considered in this study are discussed in the following subsection.

\subsection{Threshold Regression Models (TRM)}
Threshold regression models (\cite{fong2017a}) are a class of regression models in which the relationship between the predictors and the outcome variable depends on certain threshold levels, allowing the effect of predictors to change at specified time points, which are known as threshold-points or change-points.

The most general TRM, stegmented regression model (StRM) (\cite{britt2015, fong2017a}) represents a specific category of threshold regression models characterized by
\begin{equation*}
    Y_{t} = \left \{ \begin{array}{ll}
		\beta_{0} +\beta_{1}t + \varepsilon_{t} & \text{for}\ t \le t_{0}\  \\
		(\beta_{0} + \beta_{2}) + (\beta_{1} + \beta_{3})t + \varepsilon_{t} & \text{for}\ t > t_{0}\
	\end{array}
	\right.
\end{equation*}
where $\varepsilon_{t}$'s are independent and identically distributed random errors with mean 0 and variance $\sigma^{2} > 0$ and $\beta_{2}$ and $\beta_{3}$ capture the changes in intercept and slope respectively before and after the change point $t_{0}$. The model can be reformulated as
\begin{equation}
    Y_{t} = \beta_{0} + \beta_{1}t + \beta_{2}I(t > t_{0}) + \beta_{3}tI(t > t_{0}) + \varepsilon_{t}
    \label{equ:model_strm}
\end{equation}
where $I(\cdot)$ is the indicator function. This form shows the model is linear in the regression parameters when $t_{0}$ is known. In practice, however, $t_{0}$ is unknown and must be estimated, making it a central parameter of interest. Estimation of $t_{0}$ can be done by minimizing the sum of squares due to errors (SSE)
\begin{equation*}
    \sum_{t = 1}^{n} \varepsilon_{t}^{2} = \sum_{t = 1}^{t_{0}} (Y_{t} - \beta_{0} - \beta_{1} t)^2 + \sum_{t = t_{0} + 1}^{n} \{Y_{t} - (\beta_{0} + \beta_{2}) - (\beta_{1} + \beta_{3})t\}^2
\end{equation*}
with respect to $\beta_{0}, \beta_{1}, \beta_{3}, \beta_{4}$ and $t_{0}$.

A major limitation of the threshold regression model is that the sum of squared errors (SSE) is discontinuous with respect to the change point parameter $t_{0}$. Owing to this discontinuity, estimation of $t_{0}$ typically relies on a grid search over a predefined set of candidate values. However, such a grid-search approach is computationally inefficient and restricts the estimate of $t_{0}$ to a discrete set of admissible candidates, potentially limiting estimation accuracy.

\subsubsection{Segmented Regression Model (SRM)}    \label{sec:srm}
To address the discontinuity issue in StRM, the segmented regression model (\cite{fong2017b, fong2017a, liu2009, priulla2024}) ensures continuity at $t_0$ by enforcing constraints at the threshold point. The SRM is formulated as
\begin{equation}
    Y_{t} = \beta_{0} + \beta_{1}t + \beta_{2}(t - t_{0})_{+} + \varepsilon_{t}
    \label{equ:model_srm}
\end{equation}
where $(t - t_0)_{+} = \max(0, t - t_{0}) = (t - t_{0})I(t > t_{0})$. Here only slope changes before and after $t_{0}$. Thus, the segmented regression model improves upon the threshold model in two key aspects:
\begin{enumerate}[label = \roman*]
    \item it reduces the number of parameters and
    \item it eliminates the discontinuity at the change point in the error sum of squares.
\end{enumerate}

When $t_{0}$ is known, the ordinary least squares (OLS) method provides best linear unbiased estimators (BLUE) for the regression parameters and their asymptotic normality holds. However, when $t_{0}$ is unknown, as is typically the case, there is no closed-form solution for the parameter estimates. Numerical optimization techniques such as the gradient descent algorithm are not directly applicable, as the objective function involves non-differentiable elements like $I(t > t_{0})$ and $(t - t_{0})_{+}$. Consequently, derivative-free grid search based optimization methods are commonly employed. Here Binary Segmentation (BS) (\cite{scott1974, sen1975}) is employed. BS is typically a single change point method, but it can be extended to multiple change point method. To see detail of BS, see \cite{eckley2011}.

However, grid search based methods are computationally demanding, specially for large datasets. Moreover, as in the StRM framework, the grid-search approach depends critically on the predefined set of candidate values for $t_{0}$, thereby restricting the estimate to a discrete search space and reducing estimation flexibility. The model proposed in next section addresses these issues.

\section{Soft Segmented Regression Model}      \label{sec:ssrm}
As discussed in Section \ref{sec:srm}, the segmented regression model (\ref{equ:model_srm}) is more advantageous than the stegmented regression model (\ref{equ:model_strm}) in terms of both parsimony (fewer regression parameters) and continuity at the change point. Therefore, we consider the SRM (\ref{equ:model_srm}) and discuss a smooth approximation to handle its non-differentiability. The idea is to approximating the non-differentiable functions (such as $I(t > t_{0})$ or $(t - t_{0})_{+}$) with smooth, differentiable alternatives. This enables the use of efficient gradient-based optimization techniques. The idea of replacing indicator functions with smooth approximations, such as the sigmoid or the normal cumulative distribution function (CDF), can be considered as the term ``soft thresholding," as opposed to ``hard thresholding."

\subsection{Modified Segmented Regression for Change Point (MSR-CP)}     \label{sec:ssrm_detail}
In SRM (\ref{equ:model_srm}), the term $(t - t_{0})_{+}$ or equivalently the indicator function $I(t > t_0)$ introduces non-differentiability at the change point $t_{0}$, which prohibits the use of gradient-based optimization methods. To overcome the non-differentiability, our approach is to approximate the function $(t - t_{0})_{+}$ with a smooth 
sigmoid function as
\begin{equation}
    (t - t_0)_{+} = (t - t_{0}) \cdot I(t > t_{0}) \approx (t - t_0) \cdot \frac{\exp(\delta (t - t_0))}{1 + \exp(\delta (t - t_0))}    \label{equ:sigmoid}
\end{equation}
where $\delta > 0$ is a fixed tuning parameter controlling the sharpness of the transition. A larger $\delta$ yields a sharper transition, approximating the hard threshold more closely. Incorporating this approximation into SRM (\ref{equ:model_srm}), the model becomes
\begin{equation}
    Y_{t} = \beta_{0} + \beta_{1} t + \beta_{2} (t - t_{0}) \cdot \frac{\exp(\delta (t - t_{0}))}{1 + \exp(\delta (t - t_{0}))} + \varepsilon_{t}      \label{equ:model_ssrm}
\end{equation}
where $\beta_{2}$ is the difference in slope before and after the change point $t_{0}$, the tuning parameter $\delta$ controlling the sharpness of the transition and $\varepsilon_{t} \overset{i.i.d.}{\sim} WN(0, \sigma^{2})$. We refer to this formulation as the \emph{Modified Segmented Regression for Change Point (MSR-CP)}. The term involving $\delta$ acts as a smoothed version of $(t - t_{0})_{+}$, allowing the model to retain continuity and differentiability with respect to $t_{0}$.

Alternatively, we can approximate $(t - t_{0})_{+}$ directly with a smooth, differentiable maximum function, specifically the Boltzmann Operator (\cite{smooth_maximum}), as
\begin{equation*}
    (t - t_{0})_{+} \approx \frac{0 \cdot \exp(0 \cdot \delta) + (t - t_{0}) \cdot \exp(\delta (t - t_{0}))}{\exp(0 \cdot \delta) + \exp(\delta (t - t_{0}))},
\end{equation*}
which leads to the same expression obtained in (\ref{equ:sigmoid}).

However, asymptotic properties of model (\ref{equ:model_ssrm}) cannot be interpreted in the usual way. To illustrate this, we start with $n = 100$ and fix $t_{0} = 50$. Then we increase $n$ keeping $t_{0}$ fixed at 50 i.e., we consider $t_{0}$ fixed as $n$ increases. In this case, increasing the sample size ($n$) does not increase the number of observations in a fixed neighborhood of $t_{0}$. So a larger sample size does not provide any additional information about the local behavior of the model around the change point $t_{0}$. Moreover, the proportion of observations before the change point goes to 0 as $n$ becomes larger. So parameters of this regime will not be identifiable asymptotically. Alternatively, suppose we consider the change point at the middle of the time series, i.e., $t_{0} = \frac{n}{2}$. Though this setup ensures increasing number of observations in both regimes, but it also implies that $t_{0} \rightarrow \infty$ as $n \rightarrow \infty$. Thus, the change point parameter $t_{0}$ does not belong to a fixed parameter space, which complicates the formulation and interpretation of its asymptotic properties. To address both the issues, we rescale the time index of model (\ref{equ:model_ssrm}) by defining $x_{t} = \frac{t}{n}$ and $x_{0} = \frac{t_{0}}{n}$ following \cite{hillebrand2013}. Then the resulting scaled model is given by
\begin{equation}
    Y_{t} = \beta_{0} + \beta_{1} x_{t} + \beta_{2} (x_{t} - x_{0}) \cdot \frac{\exp(\delta (x_{t} - x_{0}))}{1 + \exp(\delta (x_{t} - x_{0}))} + \varepsilon_{t}      \label{equ:model_ssrm_scaled}
\end{equation}

Under this modification, the time index is rescaled from $(1, n]$ to $(0, 1]$ and the change point is represented by the normalized fixed-scale parameter $x_{0} \in (0,1]$. For example, a change point at the midpoint of the time series, $t_{0} = \frac{n}{2}$, corresponds to $x_{0} = \frac{1}{2}$ for every $n$. Furthermore, as the sample size increases, the number of observations within any fixed neighborhood of an interior point $x_{0} \in (0, 1)$ also increases. Thus, the rescaling provides an asymptotic framework in which the normalized change point remains in a fixed parameter space and an increasing amount of information becomes available in its neighborhood. This makes the scaled model (\ref{equ:model_ssrm_scaled}) suitable for studying the consistency and asymptotic distribution of the estimated change point. The mathematical justification of considering the scaled model is provided in Section \ref{sec:asympnorm_iid}. For further details, see asymptotic properties of OLS estimates of the simple time trend model in \cite{hamilton2020ch16}.

\subsection{Choice of Tuning Parameter $\delta$}    \label{sec:choice_delta}
The tuning parameter $\delta$ plays a crucial role in approximating the indicator function $I(t > t_0)$ or, equivalently, the function $(t - t_0)_+$. Specifically, the sigmoid approximation $I(t > t_0) = \frac{\exp(\delta (t - t_0))}{1 + \exp(\delta (t - t_0))}$ exhibits sharper transitions as $\delta$ increases. A larger value of $\delta$ yields a very large value of derivative in a very small interval, which results in lower variance, but with a higher bias. Conversely, smaller values of $\delta$ produces a smooth transition with lower bias and higher variance.

\begin{figure}[htbp]
    \centering
    \includegraphics{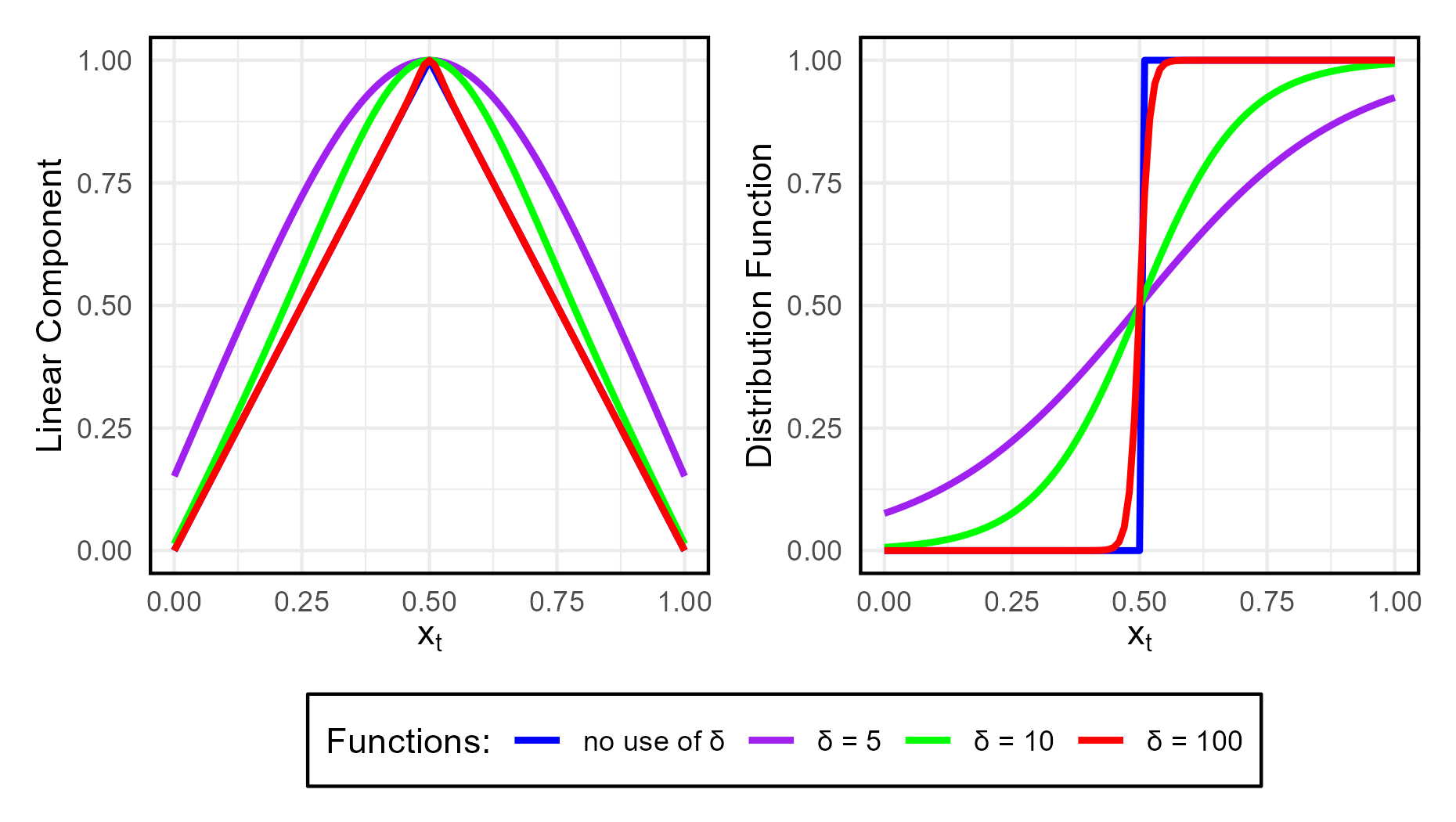}
    \caption{Comparison of sigmoid approximations with different $\delta$ values. The red line shows the true function, while the purple, green and red curves represent smooth sigmoid approximations with $\delta = 5$, $\delta = 10$ and $\delta = 100$ respectively.}
    \label{fig:sigmoid_approx}
\end{figure}

\begin{figure}
    \centering
    \includegraphics{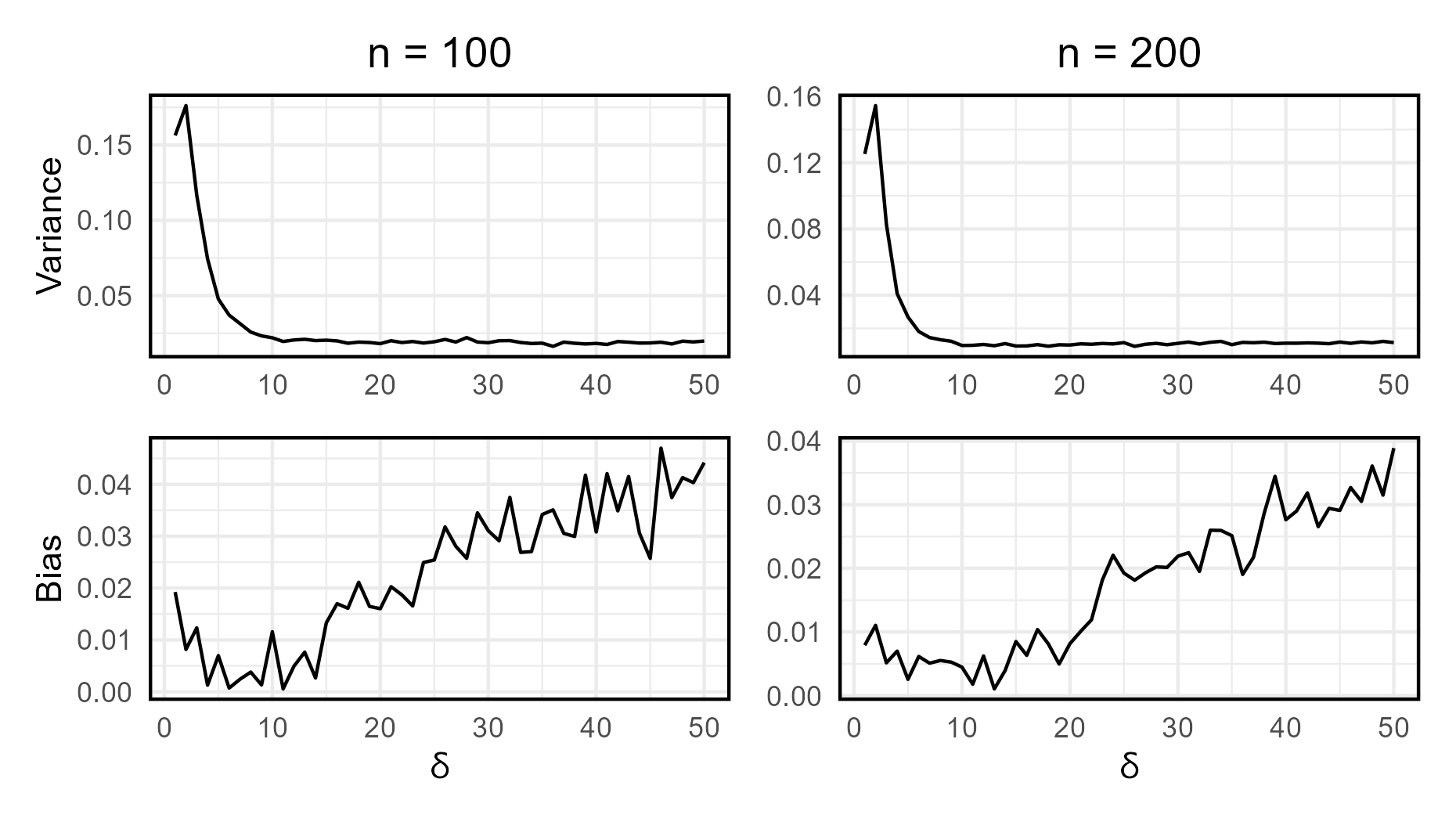}
    \caption{Bias-variance trade-off for estimated change point for different values of $\delta$}
    \label{fig:tradeoff}
\end{figure}

The smooth transition is illustrated in Figure \ref{fig:sigmoid_approx}. The blue line represents the true function $(x_{t} - x_{0})_{+}$, while the purple, green and red curves correspond to the sigmoid approximation with $\delta$ = 5, 10 and 100 respectively. As seen, increasing $\delta$ results in a closer approximation to the true function. We conduct a small simulation experiment to demonstrate the bias-variance trade-off. Figure \ref{fig:tradeoff} demonstrates this bias-variance trade-off. From this plot, the optimal value of $\delta$ may be chosen around 10, for the particular setup. For brevity, we omit the detailed simulation setup here.

We fix the value of $\delta$ separately before estimating $\boldsymbol{\theta} = (\beta_{0}, \beta_{1}, \beta_{2}, x_{0})^{\top}$. To determine an appropriate value of $\delta$, one must balance bias and variance. A simple and reasonable choice may be $\delta = \sqrt{n}$ (\cite{maiti2023}). However, for more precise estimation, here the optimal value of $\delta$ is selected using a data-driven approach. A grid search method is employed as in \cite{chakraborty2013, isl2014}. This method utilizes a goodness-of-fit measure, specifically the root mean squared error (RMSE), to calculate the optimal $\delta$. While this method is computationally intensive, it provides a principled approach to tuning parameter selection and ensures better estimation accuracy.

\subsection{Estimation of Parameters and Change Point}      \label{sec:ssrm_est}
Let $\boldsymbol{\theta} = (\beta_{0}, \beta_{1}, \beta_{2}, x_{0})^{\top}$ be the full set of parameters and $\boldsymbol{Y} = (Y_{1}, Y_{2}, \ldots, Y_{n})^{\top}$ be the vector of rainfall observations. To estimate the parameters $\beta_{0}$, $\beta_{1}$, $\beta_{2}$ and the change point $x_{0}$, we minimize the sum of squares due to errors (SSE)
\begin{equation}
    SSE(\boldsymbol{\theta}|\boldsymbol{Y}) = \sum_{t = 1}^{n} \varepsilon_{t}^{2} = \sum_{t = 1}^{n} \biggl\{Y_{t} - \beta_{0} - \beta_{1} x_{t} - \beta_{2} (x_{t} - x_{0}) \frac{\exp(\delta (x_{t} - x_{0}))}{1 + \exp(\delta (x_{t} - x_{0}))} \biggl\}^{2}     \label{equ:sse_ssrm}
\end{equation}
with respect to $\beta_{0}$, $\beta_{1}$, $\beta_{2}$ and $x_{0}$ for a chosen value of $\delta$. Note that, the SSE is nonlinear with respect to $x_{0}$, so closed form solution for the estimates of parameter is not available. But for MSR-CP model, this objective function is differentiable with respect to all parameters, including $x_{0}$, enabling the use of gradient-based optimization methods such as the Newton-Raphson or quasi-Newton algorithm.

The optimization of the objective function in (\ref{equ:sse_ssrm}) is carried out using the L-BFGS-B algorithm \cite{byrd1995} which is a quasi-Newton method and it accommodates simple box constraints on the parameters. In the present implementation, the change point parameter $x_{0}$ is restricted to the admissible range $0 < x_{0} < 1$, whereas the regression coefficients are left unconstrained. Initial values for the regression coefficients are obtained from standard linear regression fitted to a simplified model. The initial value of $x_{0}$ may be selected based on visual inspection of the time series plot or set equal to an estimate obtained from SRM or related methods. A detailed description of the estimation procedure is provided in Algorithm \ref{alg:ssrm_iid}.

\begin{algorithm}[htpb]
\caption{Change point estimation via MSR-CP under i.i.d. errors}
\label{alg:ssrm_iid}
\begin{algorithmic}[1]

\Require Time series $\{Y_{t}\}_{t = 1}^{n}$, smoothing parameter $\delta$, initial change point guess $x_0^{(0)}$
\Ensure Estimated change point $\hat{x}_{0}$, Hessian matrix $\hat{\mathcal{H}}$

\State Set $t \leftarrow (1, 2,\ldots, n)$

\State Set $x_{t} \leftarrow \frac{t}{n}$

\State Fit linear regression $Y_{t} = \alpha_{0} + \alpha_{1} x_{t} + \varepsilon_{t}$

\State Initialize parameters:
\[
\beta_{0}^{(0)} \leftarrow \hat{\alpha}_{0},\quad
\beta_{1}^{(0)} \leftarrow \hat{\alpha}_{1},\quad
\beta_{2}^{(0)} \leftarrow 0,\quad
x_{0}^{(0)} \text{ given}
\]

\State Define smooth transition function:
\[
g_{\delta}(x) = \frac{x}{1 + \exp(-\delta x)}
\]

\State Define objective function (SSE):
\[
S(\beta_{0},\beta_{1},\beta_{2},x_{0}) =
\sum_{t = 1}^{n} \Big\{Y_{t} - \beta_{0} - \beta_{1} x_{t} - \beta_{2} g_{\delta}(x_{t} - x_{0})\Big\}^{2}
\]

\State Set parameter vector $\boldsymbol{\theta} = (\beta_{0}, \beta_{1}, \beta_{2}, x_{0})^{\top}$

\State Minimize $S(\boldsymbol{\theta})$ using the L-BFGS-B algorithm subject to $0 < x_{0} < 1$.

\State Obtain optimizer $\hat{\boldsymbol{\theta}} = (\hat{\beta}_{0},\hat{\beta}_{1},\hat{\beta}_{2},\hat{x}_{0})^{\top}$

\State Compute Hessian matrix $\hat{\mathcal{H}} = \nabla^{2} S(\hat{\boldsymbol{\theta}})$

\State \Return $\hat{\beta}_{0}$, $\hat{\beta}_{1}$, $\hat{\beta}_{2}$, $\hat{x}_{0}$, $\hat{\mathcal{H}}$

\end{algorithmic}
\end{algorithm}

\begin{remark}[Choice of Initial $x_{0}$]
    The choice of the initial value $x_0$ may affect the convergence of the algorithm, potentially leading the estimate $\hat{x}_0$ to converge to a local minimum rather than the global minimum. In practice, it is advisable to run the algorithm with multiple initial values of $x_0$ and compare the resulting solutions.
\end{remark}

\begin{remark}[Extension to Multiple Change Points]
    The modified segmented model (\ref{equ:model_ssrm}) can be extended for multiple change points also. For example, the model with two change points is given by
    \begin{equation*}
        Y_{t} = \beta_{0} + \beta_{1} t + \beta_{2} (t - t_{01}) \frac{\exp(\delta_{1} (t - t_{01}))}{1 + \exp(\delta_{1} (t - t_{01}))} + \beta_{3} (t - t_{02}) \frac{\exp(\delta_{2} (t - t_{02}))}{1 + \exp(\delta_{2} (t - t_{02}))} + \varepsilon_{t}.
    \end{equation*}
    Similarly the model can be extended for multiple change points.
\end{remark}

\begin{remark}[Inclusion of External Covariate]
    Model (\ref{equ:model_ssrm}) can accommodate additional covariates such as temperature, humidity and/or other external regressors (denoted by $\boldsymbol{X_{t}}$) as follows.
    \begin{equation*}
        Y_{t} = \beta_{0} + \beta_{1} t + \beta_{2} (t - t_{0}) \cdot \frac{\exp(\delta (t - t_{0}))}{1 + \exp(\delta (t - t_{0}))} + \boldsymbol{\gamma^{\top}X_{t}} + \varepsilon_{t}
    \end{equation*}
    where $\boldsymbol{\gamma}$ is the vector of regression coefficients associated with the covariates $\boldsymbol{X_{t}}$.
\end{remark}

\subsection{Asymptotic Properties of Estimated Change Point}
Under some regularity conditions, we establish the consistency and asymptotic normality of the least square estimate of $\boldsymbol{\theta}$. Note that, minimization of objective function (\ref{equ:sse_ssrm}) is equivalent to minimizing
\begin{equation*}
    \frac{1}{n} S_{n}(\boldsymbol{\theta}) = \frac{1}{n} \sum_{t = 1}^{n} \big[ Y_{t} - f_{t}(\boldsymbol{\theta}) \big]^{2}
\end{equation*}
where $f_{t}(\boldsymbol{\theta}) = \beta_{0} + \beta_{1} x_{t} + \beta_{2} (x_{t} - x_{0}) \frac{\exp(\delta (x_{t} - x_{0}))}{1 + \exp(\delta (x_{t} - x_{0}))}$. So the least square estimate is given by
\begin{equation*}
    \boldsymbol{\hat{\theta}_{n}} = \arg \min_{\boldsymbol{\theta}} \frac{1}{n} S_{n}(\boldsymbol{\theta}).
\end{equation*}

\subsubsection{Consistency}     \label{sec:consistency_iid}
Under some regularity conditions, we establish the consistency of $\boldsymbol{\hat{\theta}_{n}}$. These conditions are listed as follows.
\begin{enumerate}[label = (A\arabic*)]
     \item \label{ass:compact_iid} The true parameter value $\boldsymbol{\theta^{*}} \in \boldsymbol{\Theta}$ where $\boldsymbol{\Theta}$ is a compact subset of $\mathbb{R}^{4}$.
     \item \label{ass:identifiability_iid} $x_{0}$ is identifiable, i.e., $\beta_{2} \ne 0$.
\end{enumerate}

These assumptions are not too restrictive from practical point of view. The boundedness assumption \ref{ass:compact_iid} is not a serious restriction because in practice, the parameters are bounded. For example, theoretically $\beta_{0}$, $\beta_{1}$ and $\beta_{2}$ can be any real number. But in practice, they will be interior points of $[-c, c]$ for some large enough $c$. For detail, see \cite{seber1989}. The change point parameter $x_{0}$ should also be a point inside the compact set $[0, 1]$, because it is not reasonable to study a change point outside of the time region of the time series. The identifiability assumption \ref{ass:identifiability_iid} is also reasonable. Because $\beta_{2} = 0$ means there is no change in slope before and after the change point and then it is immaterial to infer about the change point. This is also illustrated in the simulation study (Section \ref{sec:simulation}), where the estimation of the change point becomes inconsistent when $\beta_{2} = 0$.

\begin{theorem}[Consistency]
    Suppose that assumptions \ref{ass:compact_iid}--\ref{ass:identifiability_iid} hold. Then
    \begin{equation*}
        \boldsymbol{\hat{\theta}_{n}} \xrightarrow{p} \boldsymbol{\theta^{*}}
    \end{equation*}
    as $n \rightarrow \infty$, where ``$\xrightarrow{p}$" denotes the convergence in probability.
    \label{thm:consistency_iid}
\end{theorem}

The detailed proof of Theorem \ref{thm:consistency_iid} is provided in Appendix \ref{app:consistency_iid}. The proof is based on the consistency of nonlinear least square estimates discussed in \cite{seber1989}. Now it is worth noting that $\hat{t}_{0}$ is consistent in scaled sense.
\begin{align*}
    & \hat{x}_{0} \xrightarrow{p} x_{0}^{*} \\
    \implies & \frac{\hat{x}_{0}}{x_{0}^{*}} \xrightarrow{p} 1 \\
    \implies & \frac{\hat{t}_{0}}{t_{0}^{*}} \xrightarrow{p} 1 \\
    \implies & \forall \epsilon > 0 \quad P(|\frac{\hat{t}_{0}}{t_{0}^{*}} - 1| < \epsilon) \rightarrow 1 \text{ as } n \rightarrow \infty
\end{align*}
The statement $P(|\hat{t}_{0} - t_{0}^{*}| < \epsilon) \rightarrow 1 \text{ as } n \rightarrow \infty$ is not true in general. Rather, the ratio of estimate and true value goes to 1 in probability.

\subsubsection{Asymptotic Normality}    \label{sec:asympnorm_iid}
To establish asymptotic normality of $\boldsymbol{\hat{\theta}_{n}}$, another regularity condition is required along with \ref{ass:compact_iid} and \ref{ass:identifiability_iid}.
\begin{enumerate}[label = (A\arabic*), resume]
     \item \label{ass:interior_iid} The true parameter value $\boldsymbol{\theta^{*}}$ is an interior point of $\boldsymbol{\Theta}$.
\end{enumerate}

The regression parameters always satisfy this assumption in practice. This assumption ensures that the true change point does not occur at the boundary, i.e., $x_{0}^{*} \in (0, 1)$. This assumption is quite reasonable. Because change point cannot occur at the first or last time point of a time series.

\begin{theorem}[Asymptotic Normality]
    Suppose the assumptions \ref{ass:compact_iid}--\ref{ass:interior_iid} hold. Then
    \begin{equation*}
        \sqrt{n}(\boldsymbol{\hat{\theta}_{n}} - \boldsymbol{\theta^{*}}) \xrightarrow{d} N_{4}(\boldsymbol{0}, \sigma^{2} \Omega^{-1})
    \end{equation*}
    as $n \rightarrow \infty$, where ``$\xrightarrow{d}$" denotes the convergence in distribution and $\Omega = \lim_{n \rightarrow \infty} \frac{1}{n}\sum_{t = 1}^{n} \frac{\partial f_{t}(\boldsymbol{\theta^{*}})}{\partial \boldsymbol{\theta}} \frac{\partial f_{t}(\boldsymbol{\theta^{*}})}{\partial \boldsymbol{\theta^{\top}}}$.
    \label{thm:asympnorm_iid}
\end{theorem}

The detailed proof of Theorem \ref{thm:asympnorm_iid} is provided in Appendix \ref{app:asympnorm_iid}. The proof is based on the asymptotic normality of nonlinear least square estimates discussed in \cite{seber1989}. To prove the asymptotic normality, boundedness of $\frac{\partial f_{t}(\boldsymbol{\theta})}{\partial \boldsymbol{\theta}}$ is required for all $\boldsymbol{\theta} \in \boldsymbol{\Theta^{*}}$, an open neighborhood of $\boldsymbol{\theta^{*}}$. But if we consider the non-scaled model (\ref{equ:model_ssrm}), then $\frac{\partial f_{t}(\boldsymbol{\theta})}{\partial \beta_{1}} = t$, which is not bounded. And so the boundedness condition fails. But if the scaled model (\ref{equ:model_ssrm_scaled}) is considered, then $\frac{\partial f_{t}(\boldsymbol{\theta})}{\partial \beta_{1}} = \frac{t}{n} \le 1$ and the boundedness condition holds. This is the mathematical reason of considering the scaled model (\ref{equ:model_ssrm_scaled}) instead of the original MSR-CP model (\ref{equ:model_ssrm}), which is described in Section \ref{sec:ssrm_detail}.

\subsubsection{Estimated Asymptotic Variance and Confidence Interval}
The asymptotic variance of $\hat{x}_{0}$ can be estimated by
\begin{equation*}
    \hat{\tau}^{2} = \frac{\hat{\sigma}^{2}}{ \bigg [\frac{\partial^{2}}{\partial x_{0}^{2}}S_{n}(\boldsymbol{\theta}) \bigg ]_{\boldsymbol{\theta} = \boldsymbol{\hat{\theta}}}}
\end{equation*}
where $\boldsymbol{\hat{\theta}}$ is the estimate obtained by minimizing the objective function (\ref{equ:sse_ssrm}) and $\hat{\sigma}^{2} = \frac{S_{n}(\boldsymbol{\hat{\theta}})}{n - 4}$. Then the standard error of $\hat{x}_{0}$ is $\operatorname{SE}(\hat{x}_{0}) = \hat{\tau}$ and the $100(1 - \alpha)\%$ asymptotic confidence interval for $x_{0}$ is given by
\begin{equation*}
    (\hat{x}_{0} - z_{\frac{\alpha}{2}} \hat{\tau},\ \hat{x}_{0} + z_{\frac{\alpha}{2}} \hat{\tau}).
\end{equation*}

Now using the relation $\hat{t}_{0} = n \hat{x}_{0}$, we get $\operatorname{SE}(\hat{t}_{0}) = n \hat{\tau}$. Then $100(1 - \alpha)\%$ asymptotic confidence interval for $t_{0}$ is given by
\begin{equation*}
    (\hat{t}_{0} - z_{\frac{\alpha}{2}} \operatorname{SE}(\hat{t}_{0}),\ \hat{t}_{0} + z_{\frac{\alpha}{2}} \operatorname{SE}(\hat{t}_{0})).
\end{equation*}

\section{MSR-CP under Autoregressive Errors}      \label{sec:ssrm_arp}
In the previous section, the modified segmented reression for change point is discussed when the error are independent and identically distributed. But this i.i.d. nature of errors is very unlikely in time series data like precipitation data. In this section, we extend the estimation method for MSR-CP when the errors follow AR($p$) process, i.e., autoregressive process of order $p$. First, we write the MSR-CP model (\ref{equ:model_ssrm_scaled}) as
\begin{equation*}
    Y_{t} = \beta_{0} + \beta_{1}x_{t} + \beta_{2} z_{t}(x_{0}) + \varepsilon_{t}
\end{equation*}
where $z_{t}(x_{0}) = \frac{(x_{t} - x_{0})\exp(\delta(x_{t} - x_{0}))}{1 + \exp(\delta(x_{t} - x_{0}))}$. Now consider the error process $\{\varepsilon_{t}\}$ follows a $p^{th}$ order stationary autoregressive model:
\begin{equation}
    \varepsilon_{t} = \phi_{1} \varepsilon_{t - 1} + \phi_{2} \varepsilon_{t - 2} + \ldots + \phi_{p} \varepsilon_{t - p} + u_{t}
    \label{equ:error_arp}
\end{equation}
where $u_{t}$'s are i.i.d. errors with mean 0 and variance $\sigma^{2} > 0$. Under this specification, it is straightforward (\cite{hamilton2020}) to show that
\begin{equation*}
    \operatorname{E}(\varepsilon_{t}) = 0
\end{equation*}
and by Yule-Walker equation
\begin{equation*}
    \gamma(h) = \operatorname{Cov}(\varepsilon_{t}, \varepsilon_{t - h}) = \Bigg \{ 
    \begin{array}{ll}
        \sum_{j = 1}^{p} \phi_{j} \gamma(j) +\sigma^{2} & \text{ if } h = 0 \\
        \sum_{j = 1}^{p} \phi_{j} \gamma(h - j) & \text{ if } h = 1, 2, \ldots
    \end{array}
    \text{ with } \gamma(h) = \gamma(-h).
\end{equation*}

Now $\sum_{j = 1}^{p} \phi_{i} \gamma(h - i) = 0$ if and only if $\phi_{j} = 0$ for all $j = 1, 2, \ldots, p$, which is not the case here. So the covariance structure of the errors is non-diagonal and OLS is no longer efficient. To address this, we employ a feasible generalized least square approach based on \cite{prais1954}. The procedure by \cite{prais1954} is an extension of the Cochrane-Orcutt transformation (\cite{cochrane1949}). Before going to detail of the approach, we briefly discuss the Cochrane-Orcutt and Prais-Winsten transformation.

\subsection{Cochrane-Orcutt and Prais-Winsten Transformation}       \label{sec:ssrm_ar1}
The Cochrane-Orcutt transformation (\cite{cochrane1949}) is originally formulated for AR(1) errors. Consider $p = 1$ in the autoregressive model (\ref{equ:error_arp}) and $\phi_{1} = \phi$:
\begin{equation}
    \varepsilon_{t} = \phi \varepsilon_{t - 1} + u_{t}, \quad |\phi| < 1.
    \label{equ:error_ar1}
\end{equation}
For $t = 2, 3, \ldots, n$, the Cochrane-Orcutt transformation is given by
\begin{equation}
    Y_{t}^{*} = Y_{t} - \phi Y_{t - 1} = \beta_{0} (1 - \phi) + \beta_{1} (x_{t} - \phi x_{t - 1}) + \beta_{2} \{z_{t}(x_{0}) - \phi z_{t - 1}(x_{0})\} + (\varepsilon_{t} - \phi \varepsilon_{t - 1})
    \label{equ:transformed_Yt}
\end{equation}

This transformation neglects the first observation. On the other hand, the Prais-Winsten transformation (\cite{prais1954}) preserves the first observation in the model. This is based on the approach of feasible generalized least squares. In this transformation, in addition to the above transformation, the first observation is transformed as
\begin{equation}
    Y_{1}^{*} = \sqrt{1 - \phi^{2}} Y_{1} = \sqrt{1 - \phi^{2}} \beta_{0} + \sqrt{1 - \phi^{2}} \beta_{1} x_{1} + \sqrt{1 - \phi^{2}} \beta_{2} z_{1}(x_{0}) + \sqrt{1 - \phi^{2}} \varepsilon_{1}.
    \label{equ:transformed_Y1}
\end{equation}
Then the parameters are estimated by a iteratively reweighted least square estimation making it a special case of generalized least square. We propagate this idea for AR($p$) errors.

\subsection{Extension for AR(p) Error}      \label{sec:ssrm_arp_detail}
To develop the above transformations for AR($p$) errors, we consider the following extension of Cochrane-Orcutt transformation. For $t = p + 1, p + 2, \ldots, n$, the transformation is given by
\begin{align*}
    Y_{t}^{*} 
    &= Y_{t} - \sum_{j = 1}^{p} \phi_{j} Y_{t - j} \\
    &= \beta_{0} \big ( 1 - \sum_{j = 1}^{p} \phi_{j} \big ) + \beta_{1} \big (x_{t} - \sum_{j = 1}^{p} \phi_{j} x_{t - j} \big ) + \beta_{2} \big (z_{t}(x_{0}) - \sum_{j = 1}^{p} \phi_{j}z_{t - j}(x_{0})\big ) + \big ( \varepsilon_{t} - \sum_{j = 1}^{p} \phi_{j} \varepsilon_{t - j} \big ) \\
    &= \beta_{0} \big ( 1 - \sum_{j = 1}^{p} \phi_{j} \big ) + \beta_{1} x_{t}^{*} + \beta_{2} z_{t}^{*}(x_{0}) + u_{t}
\end{align*}
where $x_{t}^{*} = x_{t} - \sum_{j = 1}^{p} \phi_{j} x_{t - j}$ and $z_{t}^{*}(x_{0}) = z_{t}(x_{0}) - \sum_{j = 1}^{p} \phi_{j}z_{t - j}(x_{0})$. In vector notation, the transformed model is
\begin{equation}
    \boldsymbol{Y_{(p + 1 : n)}^{*}} = \beta_{0} \boldsymbol{e_{n - p}^{*}} + \beta_{1} \boldsymbol{x_{(p + 1 : n)}^{*}} + \beta_{2} \boldsymbol{z_{(p + 1 : n)}^{*}}(x_{0}) + \boldsymbol{u_{(p + 1 : n)}}
    \label{equ:model_transformed_arp_2}
\end{equation}
where $\boldsymbol{Y_{(p + 1 : n)}} = (Y_{p + 1}, Y_{p + 2}, \ldots, Y_{n})^{\top}$, $\boldsymbol{e_{n - p}^{*}} = \big ( 1 - \sum_{j = 1}^{p} \phi_{j} \big ) \boldsymbol{e_{n - p}}$, $\boldsymbol{e_{n - p}}$ is a vector of length $n - p$ with element 1, $\boldsymbol{x_{(p + 1 : n)}^{*}} = (x_{p + 1}^{*}, x_{p + 2}^{*}, \ldots, x_{n}^{*})^{\top}$, $\boldsymbol{z_{(p + 1 : n)}^{*}} = (z_{p + 1}^{*}, z_{p + 2}^{*}, \ldots, z_{n}^{*})^{\top}$ and $\boldsymbol{u_{(p + 1 : n)}} = (u_{p + 1}, u_{p + 2}, \ldots, u_{n})^{\top}$.

In Cochrane-Orcutt transformation for AR(1) errors, the first term is neglected. The Prais-Winsten transformation allows to include the first observation. Similarly, in the above transformation, first $p$ observations are not taken into account. To accommodate first $p$ observations in the model, we consider an extended Prais-Winsten transformation. The idea is to transform first $p$ observations using the Cholesky decomposition of the covariance matrix of first $p$ error terms. The procedure is as follows.

Let $\boldsymbol{\varepsilon_{(1:p)}} = (\varepsilon_{1}, \varepsilon_{2}, \ldots, \varepsilon_{p})^{\top}$ be the vector of first $p$ error terms and $\Sigma_{p}$ be the covariance matrix of the first $p$ error terms, i.e., $\Sigma_{p} = \operatorname{Cov}(\boldsymbol{\varepsilon_{(1:p)}})$. Being a variance-covariance matrix, $\Sigma_{p}$ is a symmetric positive definite matrix and so we can use Cholesky decomposition to write $\Sigma_{p} = \sigma^{2} LL^{\top}$, where $L$ is a $p \times p$ real lower triangular matrix with positive diagonal elements. Let $\boldsymbol{Y_{(1:p)}} = (Y_{1}, Y_{2}, \ldots, Y_{p})^{\top}$ be the vector of first $p$ observations. Then for $t = 1, 2, \ldots, p$, $Y_{t}$'s are transformed as
\begin{align}
    \boldsymbol{Y_{(1:p)}^{*}}
    &= L^{-1} \boldsymbol{Y_{(1:p)}} \notag \\
    &= \beta_{0} L^{-1} \boldsymbol{e_{p}} + \beta_{1} L^{-1} \boldsymbol{x_{(1:p)}} + \beta_{2} L^{-1} \boldsymbol{z_{(1:p)}}(x_{0}) + L^{-1} \boldsymbol{\varepsilon_{(1:p)}} \notag \\
    &= \beta_{0} \boldsymbol{e_{p}^{*}} + \beta_{1} \boldsymbol{x^{*}_{(1:p)}} + \beta_{2} \boldsymbol{z^{*}_{(1:p)}}(x_{0}) + \boldsymbol{\varepsilon_{(1:p)}^{*}}
    \label{equ:model_transformed_arp_1}
\end{align}
where $\boldsymbol{e_{p}}$ is a vector of $p$ 1's, $\boldsymbol{x_{(1:p)}} = (x_{1}, x_{2}, \ldots, x_{p})^{\top}$, $\boldsymbol{z_{(1:p)}}(x_{0}) = (z_{1}(x_{0}), z_{2}(x_{0}), \ldots, z_{p}(x_{0}))^{\top}$, $\boldsymbol{e_{p}^{*}} = L^{-1} \boldsymbol{{e_{p}}}$, $\boldsymbol{x_{(1:p)}^{*}} = L^{-1} \boldsymbol{x_{(1:p)}}$, $\boldsymbol{z_{(1:p)}^{*}}(x_{0}) = L^{-1} \boldsymbol{z_{(1:p)}}(x_{0})$ and $\boldsymbol{\varepsilon_{(1:p)}^{*}} = L^{-1} \boldsymbol{\varepsilon_{(1:p)}}$. Note that,
\begin{equation*}
    \operatorname{E}(\boldsymbol{\varepsilon_{(1:p)}^{*}}) = L^{-1} \operatorname{E}(\boldsymbol{\varepsilon_{(1:p)}}) = \boldsymbol{0} \text{ and}
\end{equation*}
\begin{equation*}
    \operatorname{Var}(\boldsymbol{\varepsilon_{(1:p)}^{*}}) = \operatorname{Var}(L^{-1} \boldsymbol{\varepsilon_{(1:p)}}) = L^{-1} \operatorname{Var}(\boldsymbol{\varepsilon_{(1:p)}}) (L^{-1})^{\top} = \sigma^{2} L^{-1} \Sigma_{p} (L^{-1})^{\top} = \sigma^{2} I_{p}.
\end{equation*}
So the errors of the transformed model are independent with mean 0 and variance $\sigma^{2} > 0$. Now by combining (\ref{equ:model_transformed_arp_1}) and (\ref{equ:model_transformed_arp_2}), the transformed model is written as
\begin{align*}
    \boldsymbol{Y^{*}} = 
    \begin{bmatrix}
        \boldsymbol{Y_{(1:p)}^{*}} \\
        \boldsymbol{Y_{(p + 1 : n)}^{*}}
    \end{bmatrix} &=
    \begin{bmatrix}
        \beta_{0} \boldsymbol{e_{p}^{*}} + \beta_{1} \boldsymbol{x^{*}_{(1:p)}} + \beta_{2} \boldsymbol{z^{*}_{(1:p)}}(x_{0}) \\
        \beta_{0} \boldsymbol{e_{n - p}^{*}} + \beta_{1} \boldsymbol{x_{(p + 1 : n)}^{*}} + \beta_{2} \boldsymbol{z_{(p + 1 : n)}^{*}}(x_{0})
    \end{bmatrix} +
    \begin{bmatrix}
        \boldsymbol{\varepsilon_{(1:p)}^{*}} \\
        \boldsymbol{u_{(p + 1 : n)}}
    \end{bmatrix} \\
    &= \beta_{0}
    \begin{bmatrix}
        \boldsymbol{e_{p}^{*}} \\ \boldsymbol{e_{n - p}^{*}}
    \end{bmatrix} +
    \beta_{1}
    \begin{bmatrix}
        \boldsymbol{x^{*}_{(1:p)}} \\ \boldsymbol{x^{*}_{(p + 1 : n)}}
    \end{bmatrix} +
    \beta_{2}
    \begin{bmatrix}
        \boldsymbol{z^{*}_{(1:p)}}(x_{0}) \\ \boldsymbol{z^{*}_{(p + 1 : n)}}(x_{0})
    \end{bmatrix} +
    \begin{bmatrix}
        \boldsymbol{\varepsilon_{(1:p)}^{*}} \\
        \boldsymbol{u_{(p + 1 : n)}}
    \end{bmatrix}
\end{align*}

Now writing $\boldsymbol{e^{*}} = 
\begin{bmatrix}
    \boldsymbol{e_{p}^{*}} \\ \boldsymbol{e_{n - p}^{*}}
\end{bmatrix}$, $\boldsymbol{x^{*}} = 
\begin{bmatrix}
    \boldsymbol{x^{*}_{(1:p)}} \\ \boldsymbol{x^{*}_{(p + 1 : n)}}
\end{bmatrix}$, $\boldsymbol{z^{*}}(x_{0}) =
\begin{bmatrix}
    \boldsymbol{z^{*}_{(1:p)}}(x_{0}) \\ \boldsymbol{z^{*}_{(p + 1 : n)}}(x_{0})
\end{bmatrix}$ and $\boldsymbol{u^{*}} =
\begin{bmatrix}
    \boldsymbol{\varepsilon_{(1:p)}^{*}} \\ \boldsymbol{u_{(p + 1 : n)}}
\end{bmatrix}$, it becomes
\begin{equation}
    \boldsymbol{Y^{*}} = \beta_{0} \boldsymbol{e^{*}} + \beta_{1} \boldsymbol{x^{*}} + \beta_{2} \boldsymbol{z^{*}}(x_{0}) + \boldsymbol{u^{*}}
    \label{equ:model_transformed_arp}
\end{equation}
where $\operatorname{E}(\boldsymbol{u^{*}}) = \boldsymbol{0}$ and $\operatorname{Cov}(\boldsymbol{u^{*}}) = \sigma^{2} I_{n}$. So the least square estimate is efficient for this transformed model (\ref{equ:model_transformed_arp}).

\begin{remark}
    For any other non-i.i.d. error structure, similar generalized least square technique can be used. Suppose $\operatorname{Cov}(\boldsymbol{\varepsilon}) = \Sigma$. Then $\boldsymbol{Y} = (Y_{1}, Y_{2}, \ldots, Y_{n})^{\top}$ can be transformed as $\boldsymbol{Y^{*}} = L^{-1} \boldsymbol{Y}$, where $L$ is obtained as $\Sigma = L L^{\top
    }$.
\end{remark}

\subsection{Estimation}     \label{sec:ssrm_arp_est}
For a given value of $\boldsymbol{\phi} = (\phi_{1}, \phi_{2}, \ldots, \phi_{p})^{\top}$, $\Sigma_{p}$ is obtained by solving the Yule-Walker equation for $h = 0, 1, \ldots, p$
\begin{equation*}
    \gamma(h) = \Bigg \{ 
    \begin{array}{ll}
        \sum_{j = 1}^{p} \phi_{j} \gamma(h - j) +\sigma^{2} & \text{ if } h = 0 \\
        \sum_{j = 1}^{p} \phi_{j} \gamma(h - j) & \text{ if } h = 1, 2, \ldots
    \end{array}
    \text{ with } \gamma(h) = \gamma(-h).
\end{equation*}

Now we minimize the sum of squares due to errors for the transformed model (\ref{equ:model_transformed_arp})
\begin{equation}
    S_{n}(\boldsymbol{\theta}) = \big [\boldsymbol{Y^{*}} - \beta_{0} \boldsymbol{e^{*}} + \beta_{1} \boldsymbol{x^{*}} + \beta_{2} \boldsymbol{z^{*}}(x_{0}) \big]^{\top} \big [\boldsymbol{Y^{*}} - \beta_{0} \boldsymbol{e^{*}} + \beta_{1} \boldsymbol{x^{*}} + \beta_{2} \boldsymbol{z^{*}}(x_{0}) \big]
    \label{equ:sse_arp}
\end{equation}
with respect to $\boldsymbol{\theta} = (\beta_{0}, \beta_{1}, \beta_{2}, x_{0})^{\top}$. We minimize SSE (\ref{equ:sse_arp}) by by derivative-based method (L-BFGS-B) as described in Section \ref{sec:ssrm_est}. Suppose $\boldsymbol{\hat{\theta}} = (\hat{\beta}_{0}, \hat{\beta}_{1}, \hat{\beta}_{2}, \hat{x}_{0})^{\top}$ be the minimizer thus obtained. Then calculate the residual vector $\boldsymbol{\hat{u}}^{*} = \boldsymbol{Y^{*}} - \hat{\beta}_{0} \boldsymbol{e^{*}} - \hat{\beta}_{1} \boldsymbol{x^{*}} - \hat{\beta_{2}} \boldsymbol{z^{*}}(\hat{x}_{0})$. Now to update the set of values of $\boldsymbol{\phi}$, we fit the regression of $u_{t}^{*}$ on $u_{t - 1}^{*}, u_{t - 2}^{*}, \ldots, u_{t - p}^{*}$, i.e., we fit
\begin{equation*}
    u_{t}^{*} = \phi_{1} u_{t - 1}^{*} + \phi_{2} u_{t - 2}^{*} + \ldots + \phi_{p} u_{t - p}^{*} + v_{t}, \quad t = p + 1, p + 2, \ldots, n
\end{equation*}
where $v_{t}$ are white noise. Then $\phi_{j}$'s are updated by OLS estimates of parameters of the above regression model, i.e., $\phi_{j}$'s are updated by minimizing
\begin{equation}
    \sum_{t = p + 1}^{n} v_{t}^{2} = \sum_{t = p + 1}^{n} (u_{t}^{*} - \phi_{1} u_{t - 1}^{*} - \phi_{2} u_{t - 2}^{*} - \ldots - \phi_{p} u_{t - p}^{*})^{2}.
    \label{equ:sse_arp_update}
\end{equation}

In practice, we start with an initial set of values $\boldsymbol{\phi^{(0)}} = (\phi_{1}^{(0)}, \phi_{2}^{(0)}, \ldots, \phi_{p}^{(0)})^{\top}$. Then $\boldsymbol{\hat{\theta}}$ and $\boldsymbol{\hat{\phi}}$ are updated by minimizing (\ref{equ:sse_arp_update}) and (\ref{equ:sse_arp}) iteratively. For demonstration purpose, we will consider $p$ = 1 in analysis of simulated and real data. The iterative process for AR(1) errors is described below.

\subsubsection{Estimation Procedure for p = 1}      \label{sec:ssrm_ar1_est}
Consider the setup for $p = 1$ in Section \ref{sec:ssrm_ar1}. Here $\Sigma_{1}$ will be the variance of the first observation, $Y_{1}$. The Yule-Walker equations are
\begin{align*}
    \gamma(0) &= \phi \gamma(1) + \sigma^{2} \text{ and} \\
    \gamma(1) &= \phi \gamma(0).
\end{align*}
Solving these equations, we get $\gamma(0) = \frac{\sigma^{2}}{1 - \phi^{2}} = \Sigma_{1}$ and $\gamma(1) = \frac{\phi \sigma^{2}}{1 - \phi^{2}}$. Then we write $\Sigma_{1} = \sigma^{2} L L^{\top}$ where $L = \frac{1}{\sqrt{1 - \phi^{2}}}$, considering $L$ as an $1 \times 1$ matrix. Then $Y_{1}$ is transformed as
\begin{equation*}
    Y_{1}^{*} = L^{-1} Y_{1} = \sqrt{1 - \phi^{2}} Y_{1} = \sqrt{1 - \phi^{2}} \beta_{0} + \sqrt{1 - \phi^{2}} \beta_{1} x_{1} + \sqrt{1 - \phi^{2}} \beta_{2} z_{1}(x_{0}) + \sqrt{1 - \phi^{2}} \varepsilon_{1}
\end{equation*}
which is same as obtained in (\ref{equ:transformed_Y1}). For $t = 2, 3, \ldots, n$, consider the transformation
\begin{align*}
    Y_{t}^{*} 
    &= Y_{t} - \phi Y_{t - 1} \\
    &= (\beta_{0} - \phi \beta_{0}) + \beta_{1} (x_{t} - \phi x_{t - 1}) + \beta_{2} \{z_{t}(x_{0}) - \phi z_{t - 1}(x_{0})\} + (\varepsilon_{t} - \phi \varepsilon_{t - 1}) \\
    &= \beta_{0} (1 - \phi) + \beta_{1} x_{t}^{*} + \beta_{2} z_{t}^{*}(x_{0}) + u_{t}.
\end{align*}

Now we need to minimize the objective function
\begin{equation*}
    S_{n}(\boldsymbol{\psi}) = \sum_{t = 1}^{n} \big[ Y_{t} - \beta_{0} (1 - \phi) - \beta_{1} x_{t}^{*} - \beta_{2} z_{t}^{*}(x_{0}) \big]^{2}
\end{equation*}
with respect to $\boldsymbol{\psi} = (\beta_{0}, \beta_{1}, x_{0}, \phi)^{\top}$. To perform the minimization, start with an initial $\phi = \phi^{(0)}$. Then minimize the objective function putting the value of $\phi^{(0)}$
\begin{equation*}
    S_{n}(\beta_{0}, \beta_{1}, x_{0}, \phi^{(0)}) = \sum_{t = 1}^{n} \big[ Y_{t} - \beta_{0} (1 - \phi^{(0)}) - \beta_{1} x_{t}^{*} - \beta_{2} z_{t}^{*}(x_{0}) \big]^{2}
\end{equation*}
with respect to $\beta_{0}$, $\beta_{1}$, $\beta_{2}$ and $x_{0}$ using derivative-based L-BFGS-B method as described in Section \ref{sec:ssrm_est}. Suppose $\hat{u}_{1}, \hat{u_{2}}, \ldots, \hat{u}_{n}$ are the residuals. Then $\phi^{(0)}$ is updated by minimizing
\begin{equation*}
    \sum_{t = 2}^{n}(\hat{u}_{t} - \phi \hat{u}_{t - 1})^{2}
\end{equation*}
with respect to $\phi$. This process is continued recursively until convergence. The iterative process is detailed in Algorithm \ref{alg:ssrm_ar1}.

\begin{algorithm*}[htpb]
\caption{Change point estimation via MSR-CP under AR(1) errors}
\label{alg:ssrm_ar1}
\begin{algorithmic}[1]

\Require Time series $\{Y_{t}\}_{t = 1}^{n}$, smoothing parameter $\delta$, initial change point guess $x_0^{(0)}$, tolerance $\epsilon$
\Ensure Estimated change point $\hat{t}_{0}$

\State Set $t \leftarrow (1, 2,\ldots, n)$
\State Set $x_{t} \leftarrow \frac{t}{n}$

\State Initialize $\phi^{(0)} \leftarrow 0$
\State Define transition function: $g_{\delta}(x) = \frac{x}{1 + \exp(-\delta x)}$

\Repeat

\State For $t = 1$, transform:
$Y_{1}^{*} \leftarrow \sqrt{1 - \phi^{(k)2}} Y_{1}$, $x_{1}^{*} \leftarrow \sqrt{1 - \phi^{(k)2}} x_{1}$, $g_{\delta}^{*}(x_{1}) \leftarrow \sqrt{1 - \phi^{(k)2}} g_{\delta}(x_{1})$

\State For $t = 2, \ldots, n$, transform:
$Y_{t}^{*} \leftarrow Y_{t} - \phi^{(k)} Y_{t-1}$, $x_{t}^{*} \leftarrow x_{t} - \phi^{(k)} (x_{t} - 1)$, $g_{\delta}^{*}(x_{t}) \leftarrow g_{\delta}(x_{t}) - \phi^{(k)} g_{\delta}(x_{t - 1})$

\State Estimate regression parameters and change point:

\begin{itemize}
\item Define transformed objective function: $S^{*}(\beta_{0},\beta_{1},\beta_{2},t_{0}) = \sum_{t = 1}^{n} \Big\{Y_{t}^{*} - \beta_{0}(1 - \phi^{(k)}) - \beta_{1} x_{t}^{*} - \beta_{2}g^{*}_{\delta}(x_{t} - x_{0}) \Big\}^{2}$
\item Minimize $S^{*}$ using L-BFGS-B subject to $0 < x_{0} < 1$
\item Obtain $\hat{\beta}_{0}^{(k)},\ \hat{\beta}_{1}^{(k)},\ \hat{\beta}_{2}^{(k)} \text{ and } \hat{x}_{0}^{(k)}$
\end{itemize}

\State Compute residuals: $e_{t}^{(k)} = Y_{t}^{*} - \hat{\beta}_{0}^{(k)}(1 - \phi^{(k)}) - \hat{\beta}_{1}^{(k)} x_{t}^{*} - \hat{\beta}_{2}^{(k)} g^{*}(x_{t} - \hat{x}_{0}^{(k)})$

\State Update AR(1) parameter: $\phi^{(k+1)} =
\frac{\sum_{t=2}^{n} e_{t}^{(k)} e_{t-1}^{(k)}}{\sum_{t=2}^{n} \left(e_{t}^{(k)}\right)^{2}}$

\Until{$|\phi^{(k+1)} - \phi^{(k)}| < \epsilon$}

\State Set $\hat{\beta}_{0} \leftarrow \hat{\beta}_{0}^{(k)}$, $\hat{\beta}_{1} \leftarrow \hat{\beta}_{1}^{(k)}$, $\hat{\beta}_{2} \leftarrow \hat{\beta}_{2}^{(k)}$, $\hat{\phi} \leftarrow \hat{\phi}^{(k)}$, $\hat{x}_{0} \leftarrow \hat{x}_{0}^{(k)}$

\State Compute Hessian matrix $\hat{\mathcal{H}} = \nabla^{2} S(\hat{\boldsymbol{\phi}})$

\State \Return $\hat{\beta}_{0}$, $\hat{\beta}_{1}$, $\hat{\beta}_{2}$, $\hat{\phi}$, $\hat{x}_{0}$, $\hat{\mathcal{H}}$

\end{algorithmic}
\end{algorithm*}

\subsection{Asymptotic Properties}
To study the asymptotic properties of MSR-CP model with AR($p$) errors given by (\ref{equ:error_arp}), we consider its' transformed representation given in (\ref{equ:model_transformed_arp}). The transformed errors have zero mean and a dispersion matrix whose diagonal entries are constants and off-diagonal entries are zero. So the least square estimates obtained by minimizing
\begin{equation*}
    \frac{1}{n} S_{n}(\boldsymbol{\psi}) = \frac{1}{n} \big [\boldsymbol{Y^{*}} - \beta_{0} \boldsymbol{e^{*}} + \beta_{1} \boldsymbol{x^{*}} + \beta_{2} \boldsymbol{z^{*}}(x_{0}) \big]^{\top} \big [\boldsymbol{Y^{*}} - \beta_{0} \boldsymbol{e^{*}} + \beta_{1} \boldsymbol{x^{*}} + \beta_{2} \boldsymbol{z^{*}}(x_{0}) \big]
\end{equation*}
with respect to the full parameter vector $\boldsymbol{\psi} = (\boldsymbol{\theta^{\top}}, \boldsymbol{\phi^{\top}})^{\top}$ are efficient. Suppose $\boldsymbol{\hat{\psi}_{n}}$ be the least square estimates thus obtained. Then under some suitable regularity conditions, consistency and asymptotic normality of $\boldsymbol{\hat{\psi}_{n}}$ can be established.

To establish the consistency of $\boldsymbol{\hat{\psi}_{n}}$, the following regularity conditions are required.
\begin{enumerate}[label = (A\arabic*), resume]
    \item \label{ass:compact_arp} The true parameter value $\boldsymbol{\psi^{*}} \in \boldsymbol{\Psi}$ where $\boldsymbol{\Psi}$ is a compact subset of $\mathbb{R}^{p + 4}$.
    \item \label{ass:stationarity_arp} All the roots of the equation $1 - \phi_{1}^{*} z - \phi_{2}^{*} z^{2} - \ldots - \phi_{p}^{*} z^{p} = 0$ lie outside the unit circle.
    \item \label{ass:identifiability_arp} $x_{0}$ is identifiable, i.e., $\beta_{2} \ne 0$.
\end{enumerate}

The assumption \ref{ass:stationarity_arp} is required to ensure that the true AR($p$) error process is stationary. The interpretations of the assumptions \ref{ass:compact_arp} and \ref{ass:identifiability_arp} are same as the interpretations of \ref{ass:compact_iid} and \ref{ass:identifiability_iid} described in Section \ref{sec:consistency_iid} for i.i.d. error situation.
\begin{theorem}[Consistency]
    Suppose that assumptions \ref{ass:compact_arp}--\ref{ass:identifiability_arp} hold. Then
    \begin{equation*}
        \boldsymbol{\hat{\psi}_{n}} \xrightarrow{p} \boldsymbol{\psi^{*}}
    \end{equation*}
    as $n \rightarrow \infty$, where ``$\xrightarrow{p}$" denotes the convergence in probability.
    \label{thm:consistency_arp}
\end{theorem}

Theorem \ref{thm:consistency_arp} can be proved in the similar way as that of Theorem \ref{thm:consistency_iid}. As described in Section \ref{sec:consistency_iid} for i.i.d. case, here also $\hat{t}_{0}$ is consistency in scaled sense, i.e., $\frac{\hat{t}_{0}}{t_{0}^{*}} \xrightarrow{p} 1$.

To establish asymptotic normality of $\boldsymbol{\hat{\phi}_{n}}$, consider the following additional regularity condition.
\begin{enumerate}[label = (A\arabic*), resume]
     \item \label{ass:interior_arp} The true parameter value $\boldsymbol{\psi^{*}}$ is an interior point of $\boldsymbol{\Psi}$.
\end{enumerate}
\begin{theorem}[Asymptotic Normality]
    Suppose the assumptions \ref{ass:compact_arp}--\ref{ass:interior_arp} hold. Then
    \begin{equation*}
        \sqrt{n}(\boldsymbol{\hat{\psi}_{n}} - \boldsymbol{\psi^{*}}) \xrightarrow{d} N_{p + 4}(\boldsymbol{0}, \sigma^{2} \Omega^{-1})
    \end{equation*}
    as $n \rightarrow \infty$, where ``$\xrightarrow{d}$" denotes the convergence in distribution and $\Omega = \lim_{n \rightarrow \infty} \frac{1}{n}\sum_{t = 1}^{n} \frac{\partial g_{t}(\boldsymbol{\psi^{*}})}{\partial \boldsymbol{\psi}} \frac{\partial g_{t}(\boldsymbol{\psi^{*}})}{\partial \boldsymbol{\psi^{\top}}}$, where $g_{t}(\boldsymbol{\psi})$ is the $t^{th}$ element of $\beta_{0} \boldsymbol{e^{*}} + \beta_{1} \boldsymbol{x^{*}} + \beta_{2} \boldsymbol{z^{*}}(x_{0})$.
    \label{thm:asympnorm_arp}
\end{theorem}

Theorem \ref{thm:asympnorm_arp} can be proved in the similar way as that of Theorem \ref{thm:asympnorm_iid}. The asymptotic variance of $\hat{x}_{0}$ is estimated by
\begin{equation*}
    \hat{\tau}^{2} = \frac{\hat{\sigma}^{2}}{ \bigg [\frac{\partial^{2}}{\partial x_{0}^{2}}S_{n}(\boldsymbol{\psi}) \bigg ]_{\boldsymbol{\psi} = \boldsymbol{\hat{\psi}}}}
\end{equation*}
where $\boldsymbol{\hat{\psi}}$ is the least square estimate of $\boldsymbol{\psi}$ and $\hat{\sigma}^{2} = \frac{S_{n}(\boldsymbol{\hat{\psi}})}{n - p - 4}$. Then the standard error of $\hat{x}_{0}$ is $\operatorname{SE}(\hat{x}_{0}) = \hat{\tau}$ and the standard error of $\hat{t}_{0}$, $\operatorname{SE}(\hat{t}_{0}) = n \hat{\tau}$. Then $100(1 - \alpha)\%$ asymptotic confidence interval for $t_{0}$ is given by
\begin{equation*}
    (\hat{t}_{0} - z_{\frac{\alpha}{2}} \operatorname{SE}(\hat{t}_{0}),\ \hat{t}_{0} + z_{\frac{\alpha}{2}} \operatorname{SE}(\hat{t}_{0})).
\end{equation*}

\section{Simulation Study}      \label{sec:simulation}
In this section, we conduct extensive simulation studies to demonstrate the behavior of different parameters in the MSR-CP model (\ref{equ:model_ssrm_scaled}). We generate data from the MSR-CP model (\ref{equ:model_ssrm_scaled}) with both i.i.d. and AR(1) errors. We consider different values of the parameters at 3 sample sizes ($n$) to check the performance for varying sample sizes.The small sample behavior is studied for $n = 60$, which is the sample size of our monsoon precipitation data, described in Section \ref{sec:data_examples}. The effect of moderate sample size is studied for $n = 200$, while $n = 500$ gives the large sample behavior.

At first, data of different sample sizes are generated from the MSR-CP model (\ref{equ:model_ssrm_scaled}) with i.i.d. error. To investigate the effect of each parameter on change point estimation, we consider different values of the parameter set $(\delta, \beta_{2}, x_{0})$, varying one parameter at a time while keeping the others fixed. We fix $\beta_{0} = 0$ and $\beta_{1} = 0$, as these parameters do not influence the model at the change point. The error variance is fixed at $\sigma^{2} = 4$. For each parameter setting, datasets of sizes 60, 200 and 500 are generated and the least square estimates of the parameters are obtained using the estimation procedure described in Algorithm \ref{alg:ssrm_iid}. In the estimation, the parameter $\delta$ is not estimated and is fixed at its' true value as it is a tuning parameter, as detailed in Section \ref{sec:choice_delta}. The procedure is repeated for 1000 replications. The mean and standard deviation of the least square estimates are obtained from these 1000 replications and are reported in Tables \ref{tab:sim_change_delta}--\ref{tab:sim_change_x0}. In addition, the 95\% asymptotic confidence intervals are calculated for each parameter in each replication. The coverage probability is estimated as the proportion of replications in which the corresponding true parameter value falls within its estimated confidence interval. These coverage probabilities are also reported along with the mean and standard error of the least squares estimates.

Table \ref{tab:sim_change_delta} shows the results when we vary $\delta$ keeping the other parameters fixed. The results show that in most cases the standard errors of the estimated change point decrease with increasing $\delta$. This is so because the sharpness of the transition increases with increasing $\delta$. Table \ref{tab:sim_change_beta2} reports the the results for varying $\beta_{2}$ while $\delta$ and $x_{0}$ are fixed at 40 and 0.5 respectively. Here the estimation procedure breaks completely when $\beta_{2} = 0$, due to non-identifiability of the change point parameter. Table \ref{tab:sim_change_x0} shows the results when $x_{0}$ is varied. The results indicate that the change point is estimated more accurately when the true change point is located near the middle of the time series.

Now we demonstrate the effect of varying the autoregressive parameter when the errors are AR($p$) in nature. For simplicity, we consider $p$ = 1. We generate data from MSR-CP model (\ref{equ:model_ssrm_scaled}) with AR(1) errors given in (\ref{equ:error_ar1}) for different values of $\phi$. As before, we set $\delta = 40$ and $\sigma^{2} = 4$. The least square estimates of the parameters are calculated using the estimation procedure for AR(1) errors described in Algorithm \ref{alg:ssrm_ar1}. In addition, we also estimate the parameters using Algorithm \ref{alg:ssrm_iid} to check the robustness of MSR-CP with i.i.d. errors when the data are generated using AR(1) errors. This process is replicated for 1000 times and the estimates, standard errors and the coverage probabilities are reported in Table \ref{tab:sim_change_phi}. The results indicate that the MSR-CP with AR(1) errors performs better when the true value of $\phi$ is large and positive, whereas the MSR-CP with i.i.d. errors performs well when $\phi$ is negative. When $\phi$ is close to zero, both methods exhibit comparable performance, with the MSR-CP with AR(1) errors performing slightly better, due to smaller positive value of $\phi$. In most of the above table, with increasing sample size, the bias and standard error decreases and the coverage probability goes closer to 0.95, which supports the consistency and asymptotic normality of the estimated parameters. In the next section, performance of MSR-CP in estimating change point is compared with other methods.

\begin{table}[htpb]
    \centering
    \caption{Estimates, standard errors (in parentheses) and coverage probabilities of 95\% asymptotic confidence interval (in square brackets) for MSR-CP model parameters where $\delta$ is varied}
    \begin{tabular}{|c|l|c|c|c|}
        \hline
        $\boldsymbol{\delta}$ & \textbf{Parameter} & $\boldsymbol{n = 60}$ & $\boldsymbol{n = 200}$ & $\boldsymbol{n = 500}$ \\ \hline
        \multirow{4}{*}{\rotatebox[origin = c]{90}{$\boldsymbol{\delta = 5}$}}
        & $\beta_{0}$ = 0 & $-0.31$ (1.22) [0.95] & $-0.14$ (0.70) [0.94] & $-0.04$ (0.43) [0.94] \\
        & $\beta_{1}$ = 0 & $4.28$ (11.89) [0.98] & $1.28$ (4.66) [0.94] & $0.30$ (2.21) [0.94] \\
        & $\beta_{2}$ = $-$10 & $-18.06$ (12.03) [0.99] & $-12.56$ (4.25) [0.98] & $-10.80$ (1.79) [0.98] \\
        & $x_{0}$ = 0.5 & $0.49$ (0.27) [0.86] & $0.50$ (0.17) [0.89] & $0.51$ (0.11) [0.93] \\ \hline

        \multirow{4}{*}{\rotatebox[origin = c]{90}{$\boldsymbol{\delta = 10}$}}
        & $\beta_{0}$ = 0 & $-0.16$ (0.96) [0.94] & $-0.03$ (0.47) [0.95] & $-0.01$ (0.28) [0.95] \\
        & $\beta_{1}$ = 0 & $1.14$ (6.22) [0.92] & $0.26$ (1.74) [0.95] & $0.07$ (1.03) [0.95] \\
        & $\beta_{2}$ = $-$10 & $-12.85$ (6.97) [0.95] & $-10.49$ (1.94) [0.96] & $-10.27$ (1.12) [0.96] \\
        & $x_{0}$ = 0.5 & $0.51$ (0.14) [0.87] & $0.50$ (0.07) [0.94] & $0.50$ (0.05) [0.95] \\ \hline

        \multirow{4}{*}{\rotatebox[origin = c]{90}{$\boldsymbol{\delta = 20}$}}
        & $\beta_{0}$ = 0 & $-0.06$ (0.91) [0.92] & $-0.03$ (0.42) [0.94] & $-0.02$ (0.25) [0.95] \\
        & $\beta_{1}$ = 0 & $0.36$ (6.99) [0.89] & $0.18$ (1.57) [0.92] & $0.12$ (0.87) [0.94] \\
        & $\beta_{2}$ = $-$10 & $-12.32$ (8.63) [0.94] & $-10.60$ (2.08) [0.95] & $-10.23$ (1.17) [0.95] \\
        & $x_{0}$ = 0.5 & $0.52$ (0.14) [0.78] & $0.51$ (0.07) [0.90] & $0.50$ (0.04) [0.95] \\ \hline
    \end{tabular}
    \label{tab:sim_change_delta}
\end{table}

\begin{table}[htpb]
    \centering
    \caption{Estimates, standard errors (in parentheses) and coverage probabilities of 95\% asymptotic confidence interval (in square brackets) for MSR-CP model parameters where $\beta_{2}$ is varied}
    \begin{tabular}{|l|c|c|c|}
        \hline
        \textbf{Parameter} & $\boldsymbol{n = 60}$ & $\boldsymbol{n = 200}$ & $\boldsymbol{n = 500}$ \\ \hline
        $\beta_{0}$ = 0 & $0.08$ (0.84) [0.90] & $0.00$ (0.43) [0.93] & $-0.00$ (0.26) [0.94] \\
        $\beta_{1}$ = 0 & $-0.18$ (3.30) [0.83] & $0.06$ (1.62) [0.92] & $0.03$ (0.91) [0.93] \\
        $\beta_{2}$ = $-$15 & $-17.05$ (5.74) [0.95] & $-15.21$ (1.93) [0.95] & $-15.08$ (1.25) [0.94] \\
        $x_{0}$ = 0.5 & $0.54$ (0.11) [0.73] & $0.50$ (0.05) [0.88] & $0.50$ (0.02) [0.93] \\ \hline

        $\beta_{0}$ = 0 & $-0.02$ (1.55) [0.90] & $0.01$ (0.73) [0.92] & $0.02$ (0.47) [0.93] \\
        $\beta_{1}$ = 0 & $0.63$ (21.13) [0.94] & $-0.18$ (10.36) [0.94] & $-0.28$ (7.36) [0.95] \\
        $\beta_{2}$ = 0 & $-0.33$ (22.18) [0.92] & $-0.05$ (11.36) [0.91] & $0.22$ (7.70) [0.91] \\
        $x_{0}$ = 0.5 & $0.40$ (0.19) [0.62] & $0.39$ (0.18) [0.63] & $0.37$ (0.17) [0.63] \\ \hline

        $\beta_{0}$ = 0 & $-0.03$ (0.81) [0.91] & $0.01$ (0.43) [0.94] & $0.02$ (0.26) [0.94] \\
        $\beta_{1}$ = 0 & $0.00$ (3.14) [0.85] & $-0.05$ (1.56) [0.92] & $-0.09$ (0.89) [0.94] \\
        $\beta_{2}$ = 15 & $16.70$ (4.95) [0.96] & $15.32$ (1.98) [0.96] & $15.09$ (1.22) [0.95] \\
        $x_{0}$ = 0.5 & $0.53$ (0.10) [0.76] & $0.51$ (0.05) [0.90] & $0.50$ (0.03) [0.94] \\ \hline
    \end{tabular}
    \label{tab:sim_change_beta2}
\end{table}

\begin{table}[htpb]
    \centering
    \caption{Estimates, standard errors (in parentheses) and coverage probabilities of 95\% asymptotic confidence interval (in square brackets) for MSR-CP model parameters where $x_{0}$ is varied}
    \begin{tabular}{|l|c|c|c|}
        \hline
        \textbf{Parameter} & $\boldsymbol{n = 60}$ & $\boldsymbol{n = 200}$ & $\boldsymbol{n = 500}$ \\ \hline
        $\beta_{0}$ = 0 & $0.03$ (1.93) [0.92] & $-0.07$ (0.69) [0.84] & $-0.06$ (0.46) [0.81] \\
        $\beta_{1}$ = 0 & $-2.48$ (37.49) [0.79] & $-0.14$ (7.82) [0.64] & $0.21$ (3.77) [0.74] \\
        $\beta_{2}$ = 8 & $11.72$ (37.49) [0.96] & $9.01$ (7.61) [0.88] & $8.11$ (3.50) [0.81] \\
        $x_{0}$ = 0.25 & $0.33$ (0.19) [0.58] & $0.35$ (0.17) [0.60] & $0.31$ (0.13) [0.72] \\ \hline

        $\beta_{0}$ = 0 & $0.09$ (0.96) [0.94] & $-0.01$ (0.45) [0.90] & $-0.00$ (0.26) [0.93] \\
        $\beta_{1}$ = 0 & $-0.99$ (11.19) [0.91] & $-0.06$ (1.79) [0.85] & $-0.05$ (0.98) [0.92] \\
        $\beta_{2}$ = 8 & $11.52$ (12.68) [0.94] & $8.97$ (3.24) [0.94] & $8.28$ (1.23) [0.95] \\
        $x_{0}$ = 0.5 & $0.53$ (0.16) [0.67] & $0.52$ (0.11) [0.76] & $0.51$ (0.06) [0.87] \\ \hline

        $\beta_{0}$ = 0 & $0.07$ (0.79) [0.93] & $0.06$ (0.34) [0.94] & $0.01$ (0.21) [0.94] \\
        $\beta_{1}$ = 0 & $-0.47$ (4.44) [0.95] & $-0.24$ (0.96) [0.94] & $-0.09$ (0.56) [0.93] \\
        $\beta_{2}$ = 8 & $10.68$ (15.97) [0.84] & $9.72$ (5.90) [0.92] & $8.89$ (3.50) [0.93] \\
        $x_{0}$ = 0.75 & $0.64$ (0.17) [0.66] & $0.72$ (0.12) [0.78] & $0.74$ (0.07) [0.85] \\ \hline
    \end{tabular}
    \label{tab:sim_change_x0}
\end{table}

\begin{table}[htpb]
    \centering
    \caption{Estimates, standard errors (in parentheses) and coverage probabilities of 95\% asymptotic confidence interval [in square brackets] for MSR-CP model parameters where $\phi$ is varied}
    \begin{tabular}{|l|l|c|c|c|}
        \hline
        \textbf{Parameter} & \textbf{Error} & $\boldsymbol{n = 60}$ & $\boldsymbol{n = 200}$ & $\boldsymbol{n = 500}$ \\ \hline
        \multirow{2}{*}{$\beta_{0}$ = 0} & IID & $-0.02$ (0.66) [0.98] & $-0.00$ (0.34) [0.98] & $-0.01$ (0.20) [0.99] \\
                                         & AR(1) & $-0.02$ (0.67) [0.92] & $-0.01$ (0.34) [0.91] & $-0.01$ (0.20) [0.94] \\ \hdashline
        \multirow{2}{*}{$\beta_{1}$ = 0} & IID & $0.29$ (3.12) [0.95] & $0.07$ (1.31) [0.95] & $0.09$ (0.75) [0.98] \\
                                         & AR(1) & $0.33$ (3.16) [0.84] & $0.07$ (1.31) [0.88] & $0.09$ (0.75) [0.93] \\ \hdashline
        \multirow{2}{*}{$\beta_{2}$ = $-$8} & IID & $-10.10$ (4.64) [0.99] & $-8.55$ (1.70) [0.99] & $-8.15$ (0.96) [0.99] \\
                                         & AR(1) & $-10.15$ (4.69) [0.94] & $-8.55$ (1.70) [0.94] & $-8.15$ (0.96) [0.94] \\ \hdashline
        \multirow{2}{*}{$x_{0}$ = 0.5} & IID & $0.54$ (0.14) [0.77] & $0.51$ (0.08) [0.90] & $0.50$ (0.04) [0.96] \\
                                         & AR(1) & $0.54$ (0.14) [0.67] & $0.51$ (0.08) [0.83] & $0.50$ (0.04) [0.90] \\ \hdashline
        \multirow{2}{*}{$\phi$ = $-$0.3} & IID & -- & -- & -- \\
                                         & AR(1) & $-0.33$ (0.12) [0.94] & $-0.31$ (0.06) [0.95] & $-0.30$ (0.04) [0.95] \\ \hline

        \multirow{2}{*}{$\beta_{0}$ = 0} & IID & $-0.05$ (1.00) [0.90] & $0.01$ (0.47) [0.91] & $0.00$ (0.28) [0.92] \\
                                         & AR(1) & $-0.05$ (1.00) [0.90] & $0.02$ (0.47) [0.93] & $0.00$ (0.28) [0.94] \\ \hdashline
        \multirow{2}{*}{$\beta_{1}$ = 0} & IID & $0.51$ (7.07) [0.89] & $0.01$ (1.90) [0.82] & $0.02$ (1.10) [0.88] \\
                                         & AR(1) & $0.53$ (7.08) [0.89] & $-0.01$ (2.04) [0.85] & $0.02$ (1.10) [0.91] \\ \hdashline
        \multirow{2}{*}{$\beta_{2}$ = $-$8} & IID & $-10.99$ (9.55) [0.89] & $-9.01$ (2.75) [0.92] & $-8.34$ (1.40) [0.94] \\
                                         & AR(1) & $-11.01$ (9.56) [0.91] & $-8.99$ (2.88) [0.94] & $-8.34$ (1.40) [0.96] \\ \hdashline
        \multirow{2}{*}{$x_{0}$ = 0.5} & IID & $0.52$ (0.16) [0.61] & $0.53$ (0.11) [0.72] & $0.51$ (0.07) [0.83] \\
                                         & AR(1) & $0.52$ (0.16) [0.62] & $0.53$ (0.11) [0.75] & $0.51$ (0.07) [0.86] \\ \hdashline
        \multirow{2}{*}{$\phi$ = 0.1} & IID & -- & -- & -- \\
                                         & AR(1) & $0.02$ (0.13) [0.89] & $0.08$ (0.07) [0.94] & $0.09$ (0.05) [0.94] \\ \hline

        \multirow{2}{*}{$\beta_{0}$ = 0} & IID & $-0.34$ (3.69) [0.40] & $-0.25$ (2.52) [0.42] & $-0.18$ (1.67) [0.43] \\
                                         & AR(1) & $-0.24$ (3.49) [0.70] & $-0.27$ (2.47) [0.85] & $-0.16$ (1.66) [0.91] \\ \hdashline
        \multirow{2}{*}{$\beta_{1}$ = 0} & IID & $2.37$ (22.61) [0.40] & $2.30$ (21.20) [0.43] & $0.96$ (17.25) [0.42] \\
                                         & AR(1) & $2.15$ (29.70) [0.69] & $3.15$ (32.90) [0.86] & $0.68$ (20.28) [0.92] \\ \hdashline
        \multirow{2}{*}{$\beta_{2}$ = $-$8} & IID & $-13.09$ (32.07) [0.31] & $-13.23$ (27.65) [0.35] & $-12.03$ (19.80) [0.37] \\
                                         & AR(1) & $-12.71$ (36.93) [0.61] & $-13.91$ (36.56) [0.81] & $-11.88$ (24.10) [0.89] \\ \hdashline
        \multirow{2}{*}{$x_{0}$ = 0.5} & IID & $0.47$ (0.20) [0.34] & $0.46$ (0.20) [0.27] & $0.48$ (0.19) [0.26] \\
                                         & AR(1) & $0.47$ (0.20) [0.45] & $0.46$ (0.21) [0.53] & $0.48$ (0.20) [0.58] \\ \hdashline
        \multirow{2}{*}{$\phi$ = 0.8} & IID & -- & -- & -- \\
                                         & AR(1) & $0.63$ (0.12) [0.64] & $0.75$ (0.05) [0.83] & $0.78$ (0.03) [0.90] \\ \hline
    \end{tabular}
    \label{tab:sim_change_phi}
\end{table}

\subsection{Comparison of Methods}
This section evaluates the performance of MSR-CP in estimating change point in comparison with the competing methods described in Section \ref{sec:methods}. Data are generated from the MSR-CP model (\ref{equ:model_ssrm_scaled}) with two different parameter settings:
\begin{itemize}
    \item $\boldsymbol{\beta} = (1.5, 4, -8)^{\top}$ with $x_{0} = 0.5$ and i.i.d. $N(0, 2^{2})$ errors and
    \item $\boldsymbol{\beta} = (-7, -5, 9)^{\top}$ with $x_{0} = 0.5$ and AR(1) errors with $\phi = 0.7$
\end{itemize}

\begin{figure}[htpb]
    \centering
    \includegraphics{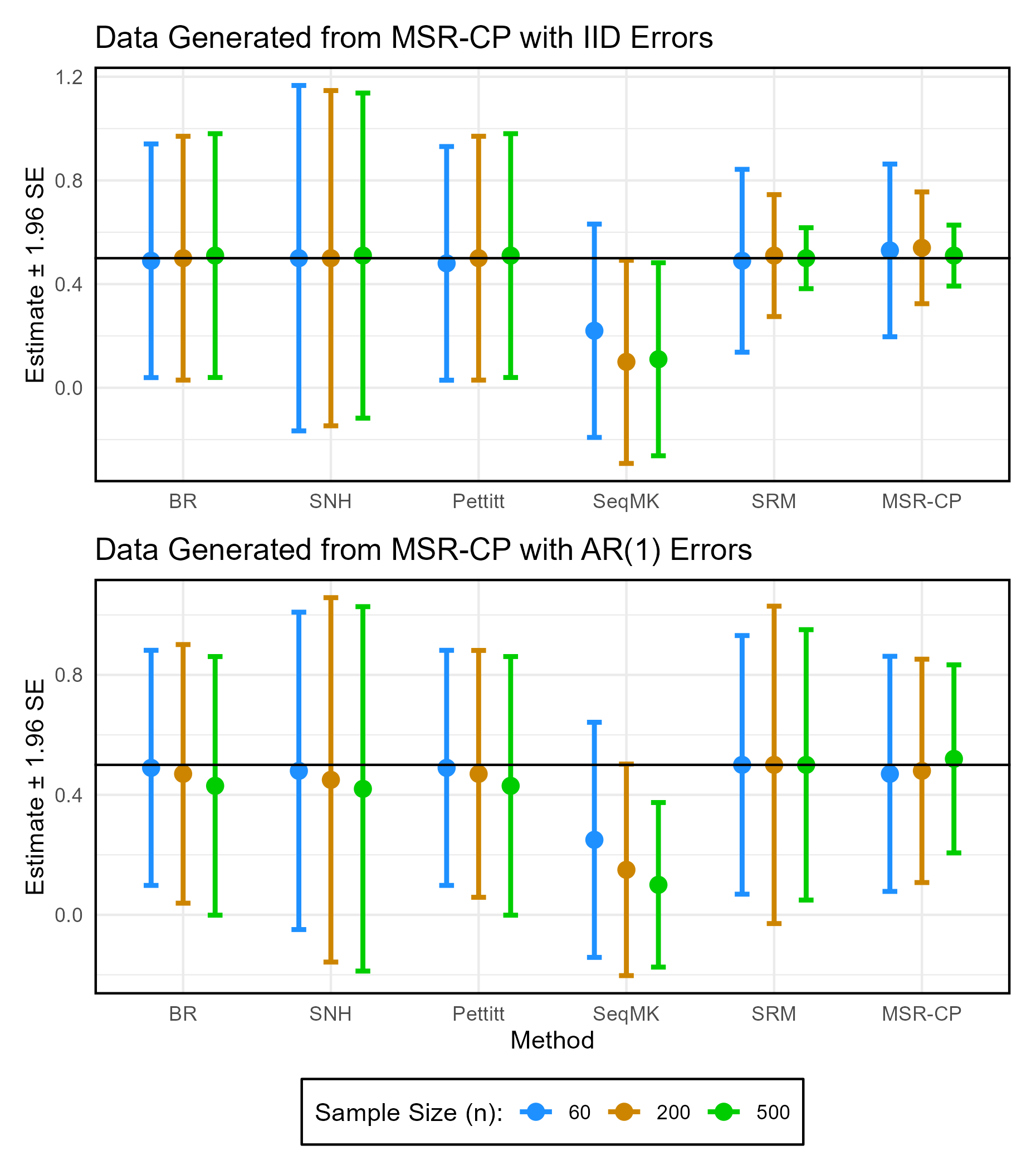}
    \caption{95\% confidence intervals for each methods when data is generated from true models}
    \label{fig:sim_comp_ssrm}
\end{figure}

In both settings, $\delta$ is fixed at 40. For each parameter settings, datasets of sizes $n$ = 60, 200 and 500 are generated and the estimates of the change point are obtained using various change point detection methods. This process is repeated for 1000 Monte Carlo replications. The mean and standard deviation of the estimated change points obtained from this 1000 replications are considered as the estimate and standard error of the change point. Bias of the estimate is calculated as the difference between estimate and true change point $x_{0} = 0.5$. The estimate, bias and standard errors are reported in Tables \ref{tab:sim_comp_ssrm_iid} and \ref{tab:sim_comp_ssrm_ar1}. The average computation time (in milliseconds) for each method is recorded for each method in the same tables. Assuming normality, the 95\% confidence interval for each method is calculated as (Estimate $\pm$ 1.96 SE) and is plotted in Figure \ref{fig:sim_comp_ssrm}.

The results indicate similar observations for both parameter settings. SRM performs best in terms of bias and standard error in estimating change point. MSR-CP follows SRM closely and the difference between SRM and MSR-CP almost vanishes especially for larger sample sizes. For both of the SRM and MSR-CP, the standard errors are decreasing with increasing sample size. MSR-CP requires much lower computation time than that of SRM, which is expected. BR and SNH exhibit very low bias but with higher standard errors and higher computation times than both of SRM and MSR-CP. Pettitt's test requires lowest computation time, attaining a very low bias but with a higher standard error even in larger sample sizes. SeqMK performs worst as it exhibits high bias across all sample sizes and the bias keep increasing with the sample size.

\begin{table}[htbp]
\footnotesize
\centering
\caption{Estimates, biases, standard errors and computation times (in milliseconds) for estimation of scaled change point when data generated from MSR-CP model with true parameters $\boldsymbol{\beta} = (1.5, 4, -8)^{\top}$ with $x_{0} = 0.5$ and i.i.d. $N(0, 2^{2})$ errors}
\begin{tabular}{|c|ccc|ccc|ccc|}
\hline
\multirow{2}{*}{\textbf{Method}}
& \multicolumn{3}{c|}{$\boldsymbol{n = 60}$} 
& \multicolumn{3}{c|}{$\boldsymbol{n = 200}$} 
& \multicolumn{3}{c|}{$\boldsymbol{n = 500}$} \\ \cline{2-10}

& Estimate & Bias (SE) & Time 
& Estimate & Bias (SE) & Time 
& Estimate & Bias (SE) & Time \\ 
\hline

BR      & 0.49 & -0.01 (0.23) & 21.86 & 0.50 & 0.00 (0.24) & 159.08 & 0.51 & 0.01 (0.24) & 997.44 \\
SNH     & 0.50 & 0.00 (0.34) & 46.46 & 0.50 & 0.00 (0.33) & 378.05 & 0.51 & 0.01 (0.32) & 2182.95 \\
Pettitt & 0.48 & -0.02 (0.23) & 0.07 & 0.50 & 0.0 (0.24) & 0.17 & 0.51 & 0.01 (0.24) & 0.69 \\
SeqMK   & 0.22 & -0.28 (0.21) & 38.17 & 0.1 & -0.36 (0.20) & 499.95 & 0.11 & -0.39 (0.19) & 3693.78 \\
SRM     & 0.49 & -0.01 (0.18) & 5.76 & 0.51 & 0.01 (0.12) & 24.01 & 0.50 & 0.00 (0.06) & 67.50 \\
MSR-CP    & 0.53 & 0.03 (0.17) & 1.30 & 0.54 & 0.04 (0.11) & 2.38 & 0.51 & 0.01 (0.06) & 5.29 \\ \hline

\end{tabular}
\label{tab:sim_comp_ssrm_iid}
\end{table}

\begin{table}[htbp]
\footnotesize
\caption{Estimates, biases, standard errors and computation times (in milliseconds) for estimation of scaled change point when data generated from MSR-CP model with true parameters $\boldsymbol{\beta} = (-7, -5, 9)^{\top}$ with $x_{0} = 0.5$ and AR(1) errors with $\phi = 0.7$}
\begin{tabular}{|c|ccc|ccc|ccc|}
\hline
\multirow{2}{*}{\textbf{Method}}
& \multicolumn{3}{c|}{$\boldsymbol{n = 60}$} 
& \multicolumn{3}{c|}{$\boldsymbol{n = 200}$} 
& \multicolumn{3}{c|}{$\boldsymbol{n = 500}$} \\ \cline{2-10}

& Estimate & Bias (SE) & Time 
& Estimate & Bias (SE) & Time 
& Estimate & Bias (SE) & Time \\ 
\hline

BR      & 0.49 & -0.01 (0.20) & 21.29 & 0.47 & -0.03 (0.22) & 159.20 & 0.43 & -0.07 (0.22) & 997.40 \\
SNH     & 0.48 & -0.02 (0.27) & 45.22 & 0.45 & -0.05 (0.31) & 378.33 & 0.42 & -0.08 (0.31) & 2182.66 \\
Pettitt & 0.49 & -0.01 (0.20) & 0.09 & 0.47 & -0.03 (0.21) & 0.19 & 0.43 & -0.07 (0.22) & 0.63 \\
SeqMK   & 0.25 & -0.25 (0.20) & 37.17 & 0.15 & -0.35 (0.18) & 506.14 & 0.10 & -0.40 (0.14) & 3686.35 \\
SRM     & 0.50 & 0.00 (0.22) & 5.55 & 0.50 & 0.00 (0.27) & 24.08 & 0.50 & 0.00 (0.23) & 67.86 \\
MSR-CP    & 0.47 & -0.03 (0.20) & 1.30 & 0.48 & -0.02 (0.19) & 2.47 & 0.52 & 0.02 (0.16) & 5.61 \\ \hline

\end{tabular}
\label{tab:sim_comp_ssrm_ar1}
\end{table}

\subsection{Robustness Check of MSR-CP}
This section investigates the robustness of MSR-CP model in estimating change point by generating data from alternative models. This allows us to examine whether the MSR-CP model maintain good estimation performance when the underlying data generating process is wrongly specified. Here we consider two alternative data generating scenarios.
\begin{itemize}
    \item \textbf{Scenario I - Data Generated from Segmented Regression Model (SRM):} In this scenario, data are generated from the segmented regression model (\ref{equ:model_srm}) discussed in Section \ref{sec:srm}. The true parameter values $\boldsymbol{\theta} = (-2, -2, 8, 0.5)$ are considered.
    \item \textbf{Scenario II - Data Generated from Probit Smoothing Model:} In this scenario, data are generated from a probit smoothing model. The probit smoothing model uses the cumulative distribution function $\Phi(\cdot)$ of Standard Normal distribution, as the sigmoid function, instead of the logistic function as in MSR-CP model. The probit smoothing model is specified as
    \begin{equation}
        Y_{t} = \beta_{0} + \beta_{1} x_{t} + \beta_{2} (x_{t} - x_{0}) \cdot \Phi(\delta (x_{t} - x_{0})) + \varepsilon_{t}
        \label{equ:model_probit}
    \end{equation}
    where the true parameter values are $\beta_{0} = 1$, $\beta_{1} = 9$, $\beta_{2} = -9$ and $x_{0} = 0.5$ with $\varepsilon_{t} \sim N(0, 2^{2})$. The value of $\delta$ is fixed at 40 as before.
\end{itemize}

\begin{figure}[htpb]
    \centering
    \includegraphics{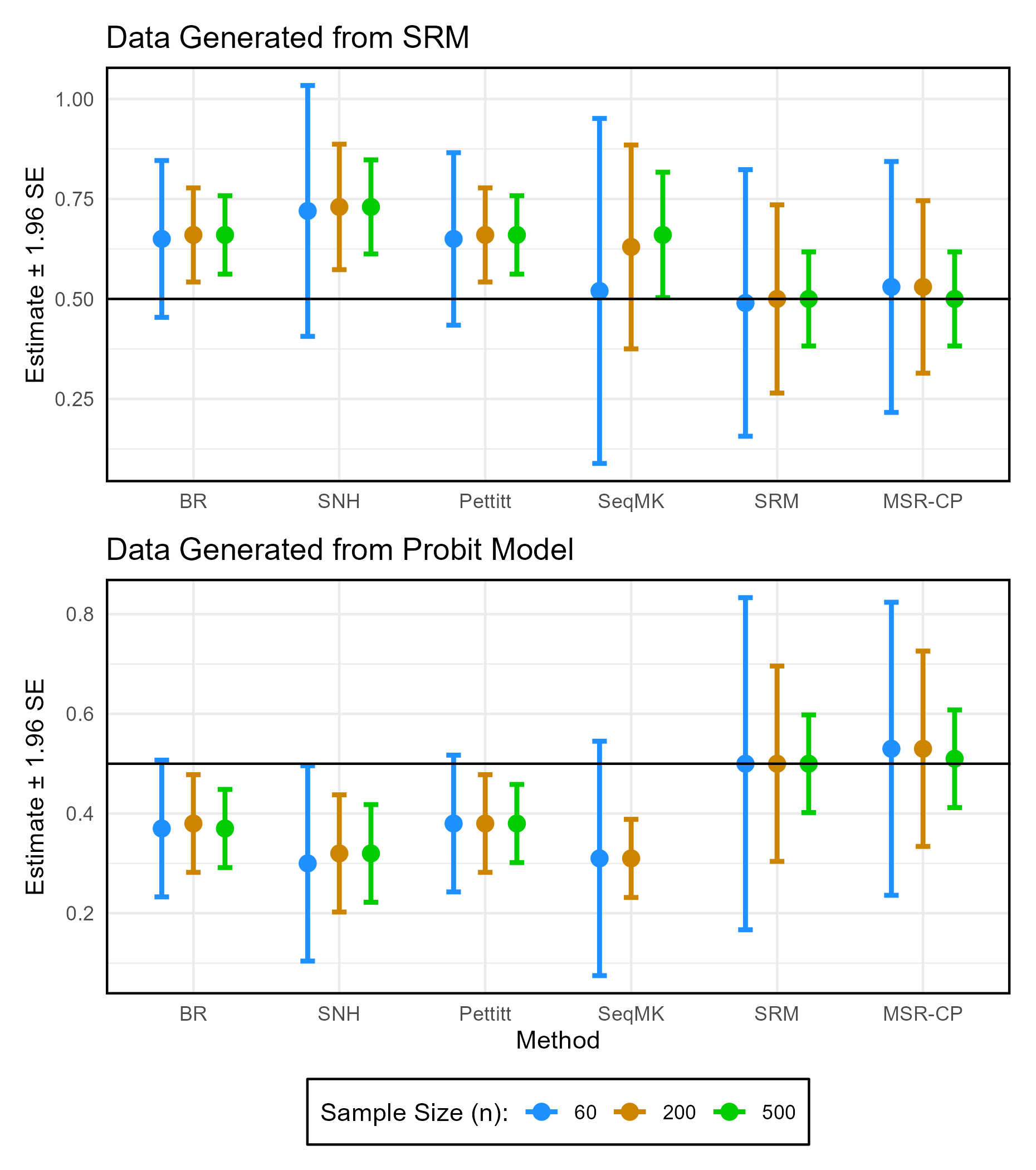}
    \caption{95\% confidence intervals for each methods when data is generated from alternative models}
    \label{fig:sim_comp_alternative}
\end{figure}

For each data generating scenario, datasets of sizes $n$ = 60, 200 and 500 are generated and the change point is estimated using various change point detection methods as in previous section. This process is repeated for 1000 Monte Carlo replications and the estimates, biases, standard errors and average computation times (in milliseconds) based on the replications are reported in Tables \ref{tab:sim_comp_srm} and \ref{tab:sim_comp_probit} respectively for above two data generating scenarios.  Assuming normality, the 95\% confidence interval for each method is calculated as (Estimate $\pm$ 1.96 SE) and is plotted in Figure \ref{fig:sim_comp_alternative}.

The results suggest similar observations for both the scenarios. SRM performs best in terms of bias and standard errors. MSR-CP performs very close to SRM but with much lower computational time. Specifically, when the sample size is large, MSR-CP achieves almost similar performance as SRM with a substantially lower computation time. None of the other methods exhibit competitive performance. SeqMK is unable to find a change point when the data are generated from probit smoothing model with $n = 500$.

\begin{table}[htbp]
\footnotesize
\caption{Estimates, biases, standard errors and computation times (in milliseconds) for estimation of scaled change point when data generated from segmented regression model (SRM)}
\begin{tabular}{|c|ccc|ccc|ccc|}
\hline
\multirow{2}{*}{\textbf{Method}}
& \multicolumn{3}{c|}{$\boldsymbol{n = 60}$} 
& \multicolumn{3}{c|}{$\boldsymbol{n = 200}$} 
& \multicolumn{3}{c|}{$\boldsymbol{n = 500}$} \\ \cline{2-10}

& Estimate & Bias (SE) & Time 
& Estimate & Bias (SE) & Time 
& Estimate & Bias (SE) & Time \\ 
\hline

BR      & 0.65 & 0.15 (0.10) & 21.14 & 0.66 & 0.16 (0.06) & 158.98 & 0.66 & 0.16 (0.05) & 997.00 \\
SNH     & 0.72 & 0.22 (0.16) & 44.92 & 0.73 & 0.23 (0.08) & 377.92 & 0.73 & 0.23 (0.06) & 2181.63 \\
Pettitt & 0.65 & 0.15 (0.11) & 0.07 & 0.66 & 0.16 (0.06) & 0.17 & 0.66 & 0.16 (0.05) & 0.55 \\
SeqMK   & 0.52 & 0.02 (0.22) & 37.14 & 0.63 & 0.13 (0.13) & 507.60 & 0.66 & 0.16 (0.08) & 3707.94 \\
SRM     & 0.49 & -0.01 (0.17) & 5.54 & 0.50 & 0.00 (0.12) & 24.00 & 0.50 & 0.00 (0.06) & 67.42 \\
MSR-CP    & 0.53 & 0.03 (0.16) & 1.24 & 0.53 & 0.03 (0.11) & 2.37 & 0.50 & 0.00 (0.06) & 5.32 \\ \hline

\end{tabular}
\label{tab:sim_comp_srm}
\end{table}

\begin{table}[htbp]
\footnotesize
\centering
\caption{Estimates, biases, standard errors and computation times (in milliseconds) for estimation of scaled change point when data generated from probit smoothing model (\ref{equ:model_probit})}
\begin{tabular}{|c|ccc|ccc|ccc|}
\hline
\multirow{2}{*}{\textbf{Method}}
& \multicolumn{3}{c|}{$\boldsymbol{n = 60}$} 
& \multicolumn{3}{c|}{$\boldsymbol{n = 200}$} 
& \multicolumn{3}{c|}{$\boldsymbol{n = 500}$} \\ \cline{2-10}

& Estimate & Bias (SE) & Time 
& Estimate & Bias (SE) & Time 
& Estimate & Bias (SE) & Time \\ 
\hline

BR      & 0.37 & -0.13 (0.07) & 21.14 & 0.38 & -0.12 (0.05) & 158.82 & 0.37 & -0.13 (0.04) & 996.85 \\
SNH     & 0.30 & -0.20 (0.10) & 44.85 & 0.32 & -0.18 (0.06) & 377.57 & 0.32 & -0.18 (0.05) & 2180.23 \\
Pettitt & 0.38 & -0.12 (0.07) & 0.07 & 0.38 & -0.12 (0.05) & 0.17 & 0.38 & -0.12 (0.04) & 0.52 \\
SeqMK   & 0.31 & -0.19 (0.12) & 37.02 & 0.31 & -0.19 (0.04) & 499.36 & -- & -- & 3688.95 \\
SRM     & 0.50 & 0.00 (0.17) & 5.53 & 0.50 & 0.00 (0.10) & 23.93 & 0.50 & 0.00 (0.05) & 67.81 \\
MSR-CP    & 0.53 & 0.03 (0.15) & 1.24 & 0.53 & 0.03 (0.10) & 2.44 & 0.51 & 0.01 (0.05) & 5.29 \\ \hline

\end{tabular}
\label{tab:sim_comp_probit}
\end{table}

\section{Data Analysis}     \label{sec:data_analysis}
\subsection{Change Point Analysis: Precipitation Data of South 24 Parganas}
In this section, we estimate the change point in the monsoon precipitation data for the South 24 Parganas district using various change point detection methods. For this dataset, the sample size ($n$) is 60. We first estimate the change point using the MSR-CP model (\ref{equ:model_ssrm_scaled}).

\begin{figure}[htpb]
    \centering
    \includegraphics{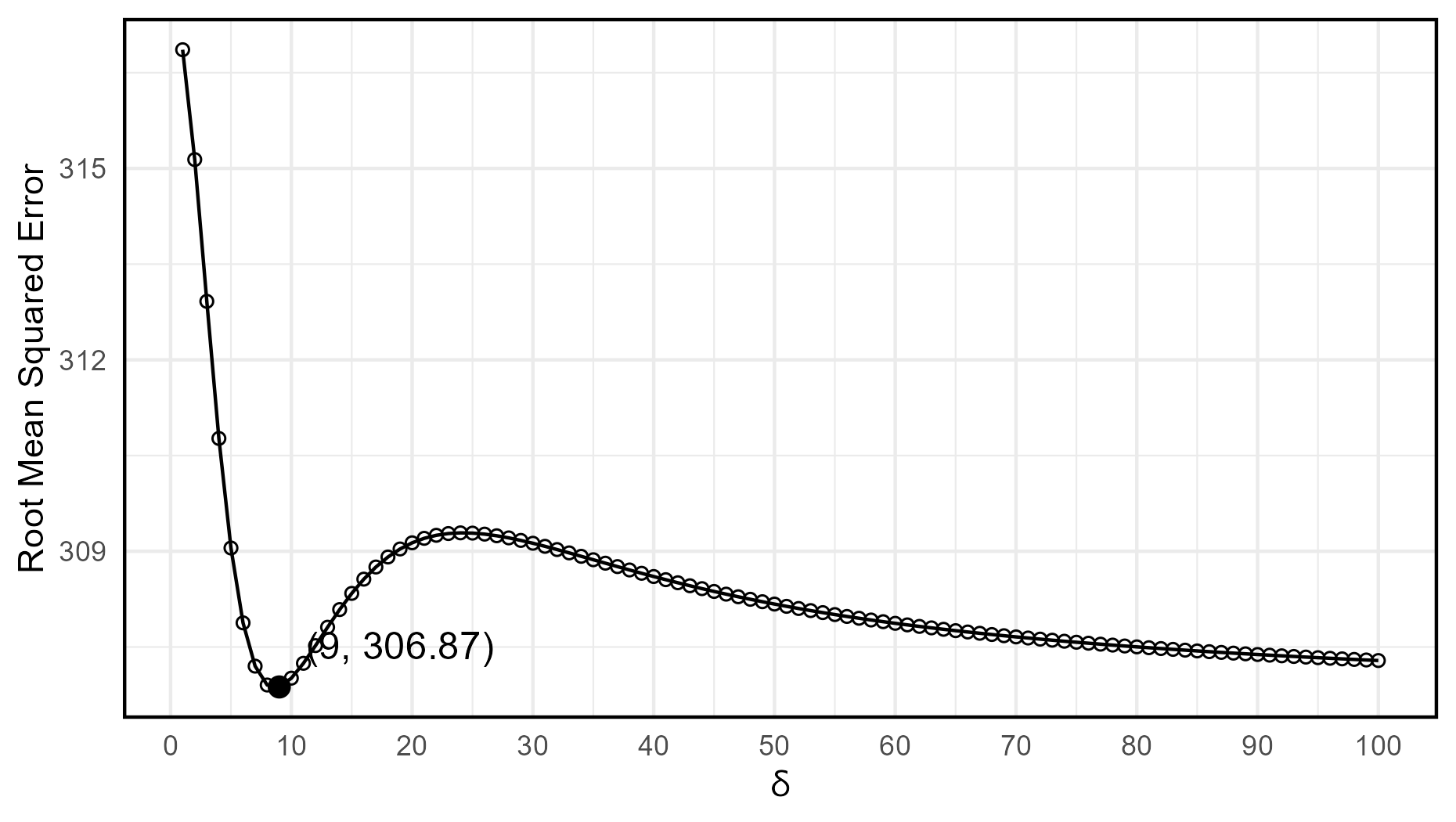}
    \caption{RMSE vs $\delta$ (for South 24 Parganas)}
    \label{fig:rmse_plot_s24pgs}
\end{figure}

To identify the change point using MSR-CP, we determine the optimal value of $\delta$. To this end, we conduct a grid search over $\delta \in \{ 1, 2, \ldots, 100 \}$ and calculate the corresponding root mean square error (RMSE) for each value. We then choose the value of $\delta$, which corresponds to the minimum RMSE. The RMSE values for different $\delta$'s are plotted in Figure \ref{fig:rmse_plot_s24pgs}. From the plot, we obtain the minimum RMSE as 306.87 and the corresponding optimum value is $\delta = 9$.

\begin{figure}[htpb]
    \centering
    \includegraphics{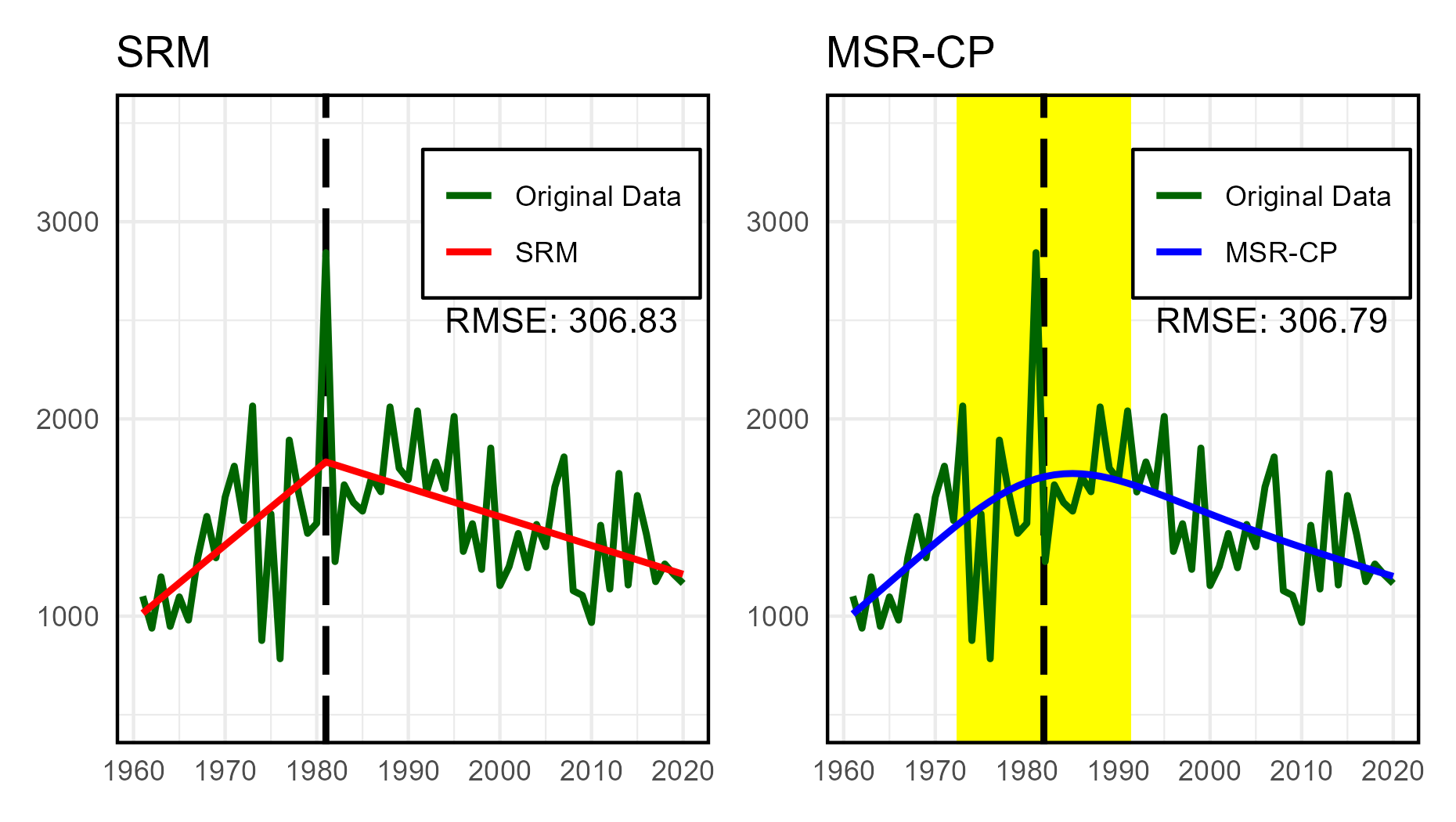}
    \caption{Fitted threshold regression models (SRM and MSR-CP) for change point estimation (for South 24 Parganas). The yellow shaded region deonotes the 95\% asymptotic confidence interval for the change point.}
    \label{fig:fitting_plot_s24pgs}
\end{figure}

To find the change point using MSR-CP (\ref{equ:model_ssrm_scaled}), we fix the value of $\delta$ at 9. Then we fit the scaled MSR-CP model (\ref{equ:model_ssrm_scaled}) with i.i.d. errors and the results are reported in Table \ref{tab:ssrm_iid_s24pgs}. The estimated change point is $\hat{t}_{0} = 21.86$ on the time-point scale. Since the time series begins in 1961, with the first observation indexed by $t = 1$, this estimate corresponds approximately to the $22^{\text{nd}}$ observation, i.e., the year 1982. The slope of the regression line before the estimated change point (1982) is obtained as 2124.04 and the corresponding after the change point is $(2124.04 - 2927.63)  = -803.59$. The 95\% asymptotic confidence interval is estimated as $(1972.30, 1991.41) \approx (1972, 1991)$, expressed in calender-year scale. The confidence interval and the fitted model is plotted in the right panel of Figure \ref{fig:fitting_plot_s24pgs}. The $p$-value for $H_{0}: \beta_{2} = 0$ is obtained as $P(Z > |\frac{\hat{\beta}_{2}}{\operatorname{SE}(\hat{\beta}_{2})}|) = 0.0002$, where $Z \sim N(0, 1)$. This provides strong evidence for existence of a change in trend in monsoon rainfall in South 24 Parganas at 1982.

\begin{figure}[htpb]
    \centering
    \includegraphics{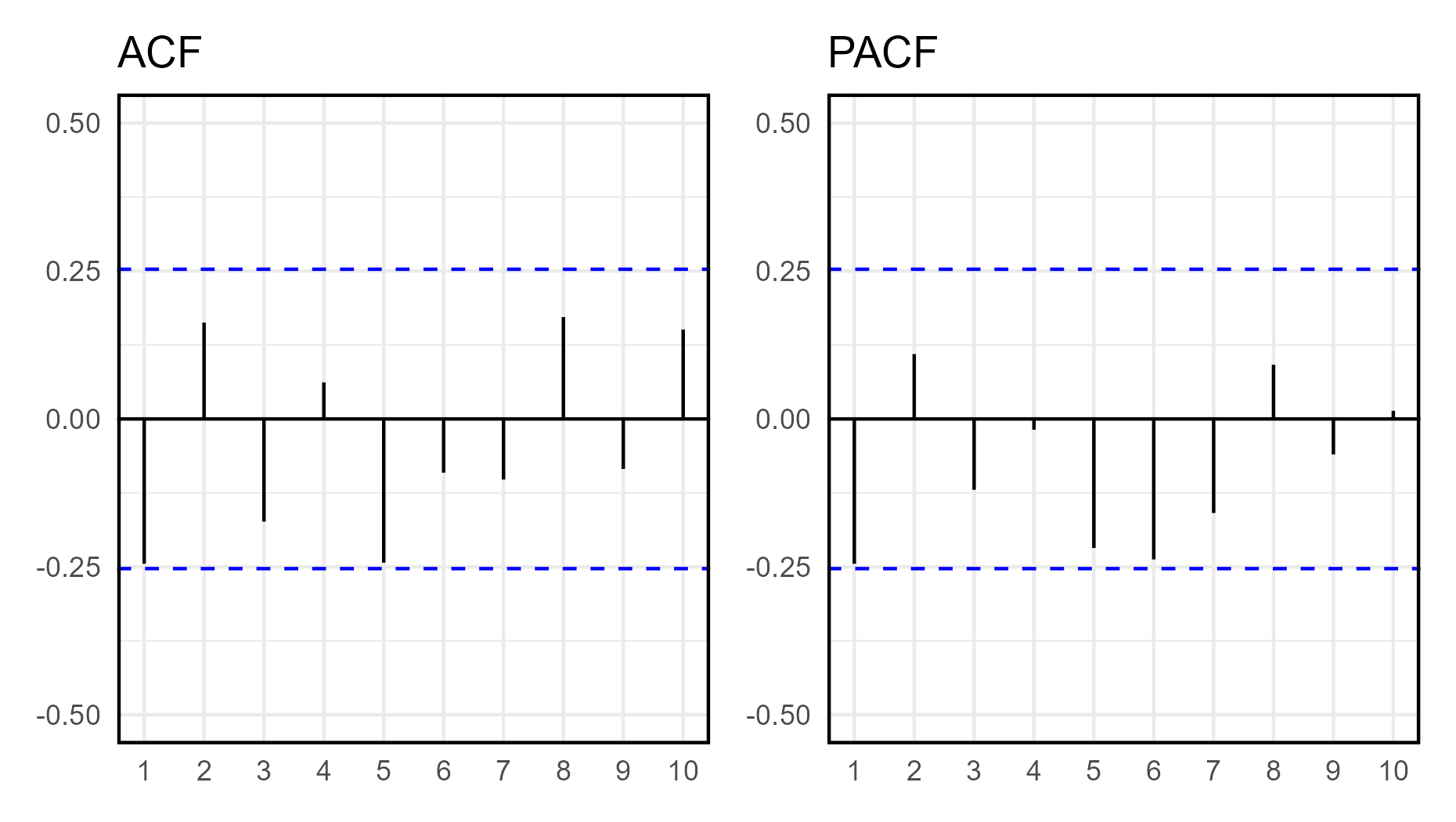}
    \caption{ACF and PACF plots for residuals obtained from MSR-CP with i.i.d. errors (for South 24 Parganas)}
    \label{fig:acfpacf_residual_s24pgs}
\end{figure}

We next examine whether the errors can reasonably be assumed to be i.i.d. To this end, we plot the autocorrelation function (ACF) and partial autocorrelation function (PACF) of the residuals obtained from the MSR-CP model fitted under the i.i.d. error assumption (Figure \ref{fig:acfpacf_residual_s24pgs}). Although neither plot provides statistically significant evidence of autocorrelation at the 5\% significance level, the first-order autocorrelation and partial autocorrelation coefficients lie close to the boundaries of the 95\% confidence intervals. This suggests that a first-order autoregressive structure is worth investigating. Accordingly, we proceed to fit the MSR-CP model with AR(1) errors to examine whether accounting for serial dependence affects the estimated change point and the associated inference.

Now we fit the MSR-CP model (\ref{equ:model_ssrm_scaled}) considering AR(1) errors, specified in (\ref{equ:error_ar1}), and estimate the change point, model parameters and the autoregressive parameter using Algorithm \ref{alg:ssrm_ar1}. The results are provided in Table \ref{tab:ssrm_ar1_s24pgs}. The change point is estimated as $\hat{t}_{0} = 21.87 \approx 22$ on the time-point scale, which is effectively same as obtained in the case of i.i.d. errors. So the change point is estimated as the year 1982 with 95\% asymptotic confidence interval $(1974, 1989)$. The estimate of the AR(1) parameter is obtained as $\hat{\phi} = -0.24$ and the $p$-value for testing $H_{0}: \phi = 0$ is $P(Z > |\frac{\hat{\phi}}{\operatorname{SE}(\hat{\phi})}|) = 0.05$.

Apart from MSR-CP, other methods of estimating change point (described in Section \ref{sec:methods}) are applied on the monsoon rainfall data of South 24 Parganas. The estimated change points and computation times obtained through these methods are presented in Table \ref{tab:cpt_s24pgs} along with change points obtained through MSR-CP. The estimates obtained from parametric (BR and SNH) and nonparametric (Pettitt's and SeqMK) approaches are not very close, because these approaches detect change in mean. But in the precipitation data of South 24 Paraganas, the change is in trend. The threshold regression based models (SRM and MSR-CP) find almost same change point. The derivative-based MSR-CP with i.i.d. errors reduces computation time almost 54\% relative to grid-search based SRM. Though the MSR-CP with AR(1) error has larger computation time than SRM, the computation time is lower than other methods except Pettitt's test. But from simulation study, we see that Pettitt's test does not provide reliable estimates of the change point when the change is in trend.

\begin{table}[htpb]
    \centering
    \caption{Estimates and standard errors of model parameters along with 95\% asymptotic confidence interval of the change point for MSR-CP with i.i.d. errors (for South 24 Parganas)}
    \begin{tabular}{|c|cc|}
        \hline
        \textbf{Parameter} & \textbf{Estimate} & \textbf{SE} \\ \hline
        $\beta_{0}$ & 932.74 & 178.45 \\
        $\beta_{1}$ & 2124.04 & 854.21 \\
        $\beta_{2}$ & $-2927.63$ & 778.41 \\
        $x_{0}$     & 0.36 & 0.08 \\
        $t_{0} (= n x_{0})$ & 21.86 & 4.87 \\ \hline
        95\% CI of $t_{0}$ & \multicolumn{2}{c|}{(12.30, 31.41)} \\ \hline
    \end{tabular}
    \label{tab:ssrm_iid_s24pgs}
\end{table}

\begin{table}[htpb]
    \centering
    \caption{Estimates and standard errors of model parameters along with 95\% asymptotic confidence interval of the change point for MSR-CP with AR(1) errors (for South 24 Parganas)}
    \begin{tabular}{|c|cc|}
        \hline
        \textbf{Parameter} & \textbf{Estimate} & \textbf{SE} \\ \hline
        $\beta_{0}$ & 931.63 & 143.26 \\
        $\beta_{1}$ & 2124.82 & 681.41 \\
        $\beta_{2}$ & $-$2927.16 & 620.82 \\
        $\phi$      & $-$0.24 & 0.13 \\
        $x_{0}$     & 0.36 & 0.06 \\
        $t_{0} (= n x_{0})$ & 21.87 & 3.85 \\ \hline
        95\% CI of $t_{0}$ & \multicolumn{2}{c|}{(14.31, 29.42)} \\ \hline
    \end{tabular}
    \label{tab:ssrm_ar1_s24pgs}
\end{table}

\begin{table}[htpb]
    \centering
    \caption{Estimated change point and computation time (in milliseconds) obtained through different methods for monsoon rainfall in South 24 Parganas}
    \begin{tabular}{|c|c|c|}
        \hline
        \textbf{Method} & \textbf{Estimated Change Point} & \textbf{Computation Time} \\ \hline
        BR & 1976 & 45.05 \\
        SNH & 1966 & 92.89 \\
        Pettitt's & 1969 & 0.18 \\
        SeqMK & 1974 & 84.90 \\
        SRM & 1981 & 12.03 \\
        MSR-CP (IID Error) & 1982 & 5.55 \\
        MSR-CP (AR(1) Error) & 1982 & 65.33 \\ \hline
    \end{tabular}
    \label{tab:cpt_s24pgs}
\end{table}

\subsection{Change Point Analysis: Precipitation Data of Coochbehar}
\begin{figure}[htpb]
    \centering
    \includegraphics{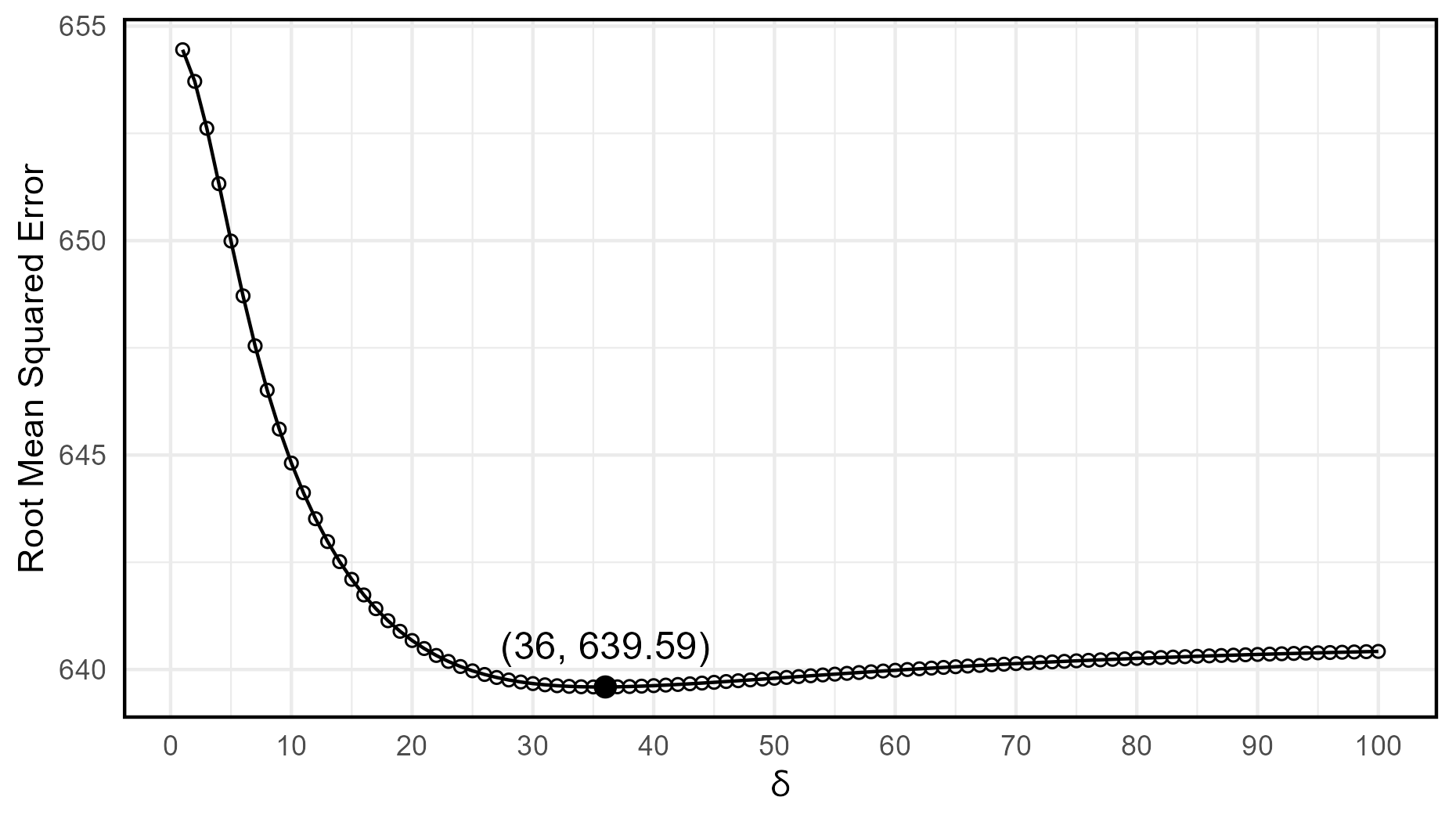}
    \caption{RMSE vs $\delta$ (for Coochbehar)}
    \label{fig:rmse_plot_coochbehar}
\end{figure}

In this section, we estimate change point in monsoon precipitation data of Coochbehar district through various methods. The sample size ($n$) is 60 for this dataset. First, we estimate the change point using MSR-CP model (\ref{equ:model_ssrm_scaled}). The RMSE vs $\delta$ plot, in Figure \ref{fig:rmse_plot_coochbehar}, indicates the optimum value of $\delta = 36$ and the corresponding RMSE is 639.59.

\begin{figure}[htpb]
    \centering
    \includegraphics{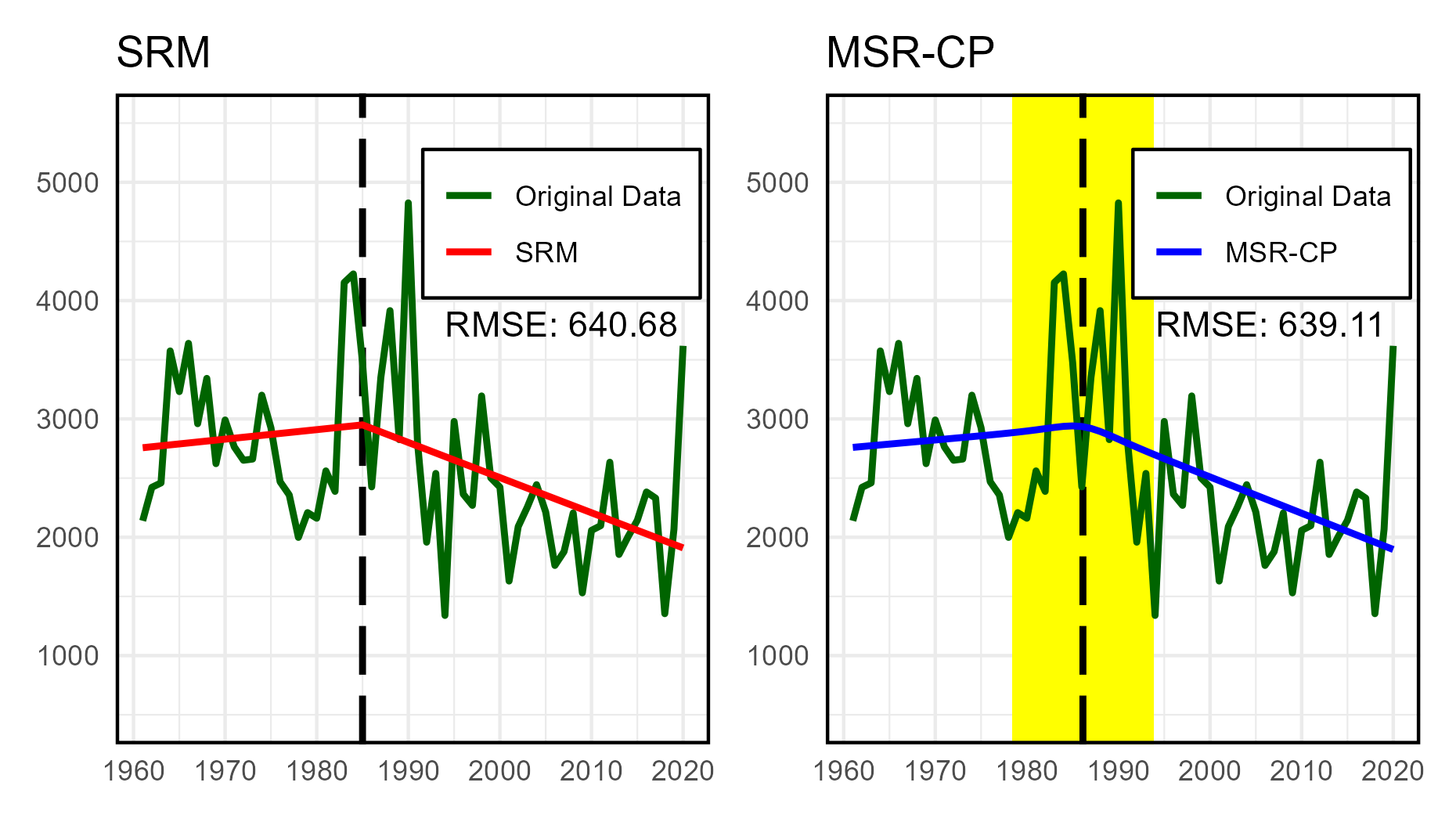}
    \caption{Fitted threshold regression models (SRM and MSR-CP) for change point estimation (for Coochbehar). The yellow shaded region deonotes the 95\% asymptotic confidence interval for the change point.}
    \label{fig:fitting_plot_coochbehar}
\end{figure}

Fixing $\delta = 36$, the results of fitting MSR-CP model (\ref{equ:model_ssrm_scaled}) with i.i.d. errors are presented in Table \ref{tab:ssrm_iid_coochbehar}. The estimated change point is $\hat{t}_{0} = 26.12 \approx 26$ on the time-point scale, which corresponds the year 1986. The slope of the regression line, representing trend in the precipitation data, before the estimated change point (1986) is obtained as 418.14 and the corresponding after the change point is $(418.14 - 2256.38)  = -1838.24$. The 95\% asymptotic confidence interval is estimated as $(1978.41, 1993.83) \approx (1978, 1994)$. The confidence interval and the fitted model is plotted in the right panel of Figure \ref{fig:fitting_plot_coochbehar}. The $p$-value for $H_{0}: \beta_{2} = 0$ is obtained as $P(Z > |\frac{\hat{\beta}_{2}}{\operatorname{SE}(\hat{\beta}_{2})}|) = 0.05$, where $Z \sim N(0, 1)$.

\begin{figure}[htpb]
    \centering
    \includegraphics{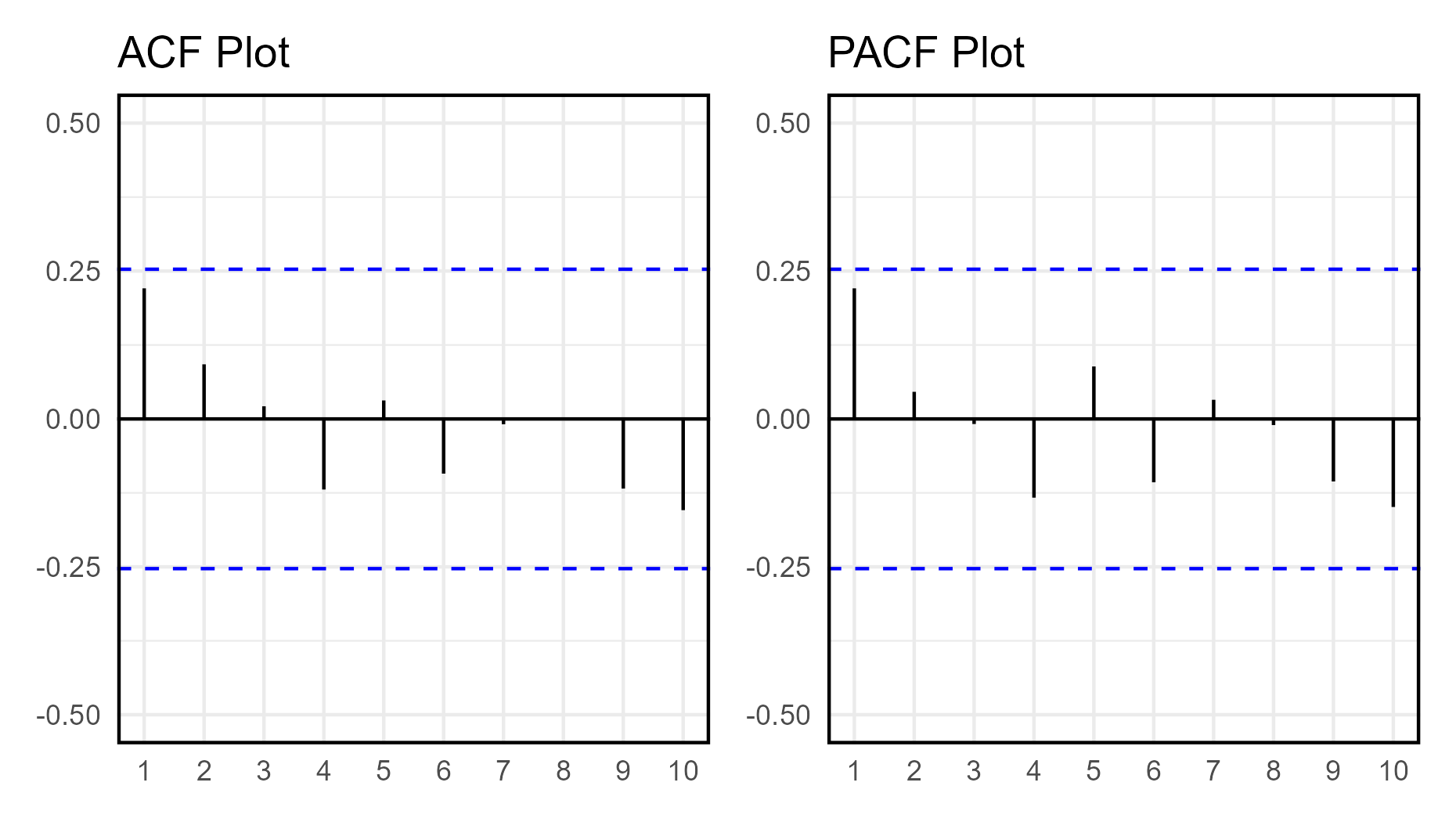}
    \caption{ACF and PACF plots for residuals obtained from MSR-CP with i.i.d. errors (for Coochbehar)}
    \label{fig:acfpacf_residual_coochbehar}
\end{figure}

The ACF and PACF are plotted for the residuals obtained from above MSR-CP model in Figure \ref{fig:acfpacf_residual_coochbehar}. As shown in the plot, the first-order autocorrelation and partial autocorrelation coefficients are very close to the boundary of the 95\% confidence interval. So we investigate the MSR-CP with AR(1) errors. Towards this end, the model parameters, AR(1) parameter and the change point are estimated using Algorithm \ref{alg:ssrm_ar1}. The results are provided in Table \ref{tab:ssrm_ar1_coochbehar}. Results show that $\hat{t}_{0} = 25.78$. That means, the estimated change point is $25.78 \approx 26^{\text{th}}$ time point, i.e., the year 1986, which is same as obtained from MSR-CP with i.i.d. errors. Also the AR(1) parameter $\phi$ is estimated as $\hat{\phi} = 0.26$. The $p$-value for testing $H_{0}: \phi = 0$ is $P(Z > |\frac{\hat{\phi}}{\operatorname{SE}(\hat{\phi})}|) = 0.04$, which shows evidence for presence of first-order autocorrelation in the data.

The estimated change points obtained from competing methods and their computation times are reported in Table \ref{tab:cpt_coochbehar}. The estimated change point obtained from BR, SNH and Pettitt's test is same, i.e., 1991. Sequential Mann-Kendall test is unable to detect any change point. The threshold regression based models (i.e., SRM and MSR-CP) find almost same change point. As expected, MSR-CP reduces computation time almost 51\% relative to that of SRM. Although the MSR-CP with AR(1) errors requires more computation time than the SRM, its computation time remains lower than that of the other methods, except Pettitt's test. However, simulation study indicates that Pettitt's test may not provide reliable change point estimates when the change occurs in the trend.

\begin{table}[htpb]
    \centering
    \caption{Estimates and standard errors of model parameters along with 95\% asymptotic confidence interval of the change point for MSR-CP with i.i.d. errors (for Coochbehar)}
    \begin{tabular}{|c|cc|}
        \hline
        \textbf{Parameter} & \textbf{Estimate} & \textbf{SE} \\ \hline
        $\beta_{0}$ & 2752.57 & 244.63 \\
        $\beta_{1}$ & 418.14 & 822.36 \\
        $\beta_{2}$ & $-$2256.38 & 1170.76 \\
        $x_{0}$     & 0.44 & 0.07 \\
        $t_{0} (= n x_{0})$ & 26.12 & 3.93 \\ \hline
        95\% CI of $t_{0}$ & \multicolumn{2}{c|}{(18.41, 33.83)} \\ \hline
    \end{tabular}
    \label{tab:ssrm_iid_coochbehar}
\end{table}

\begin{table}[htpb]
    \centering
    \caption{Estimates and standard errors of model parameters along with 95\% asymptotic confidence interval of the change point for MSR-CP with AR(1) errors (for Coochbehar)}
    \begin{tabular}{|c|cc|}
        \hline
        \textbf{Parameter} & \textbf{Estimate} & \textbf{SE} \\ \hline
        $\beta_{0}$ & 2723.36 & 317.08 \\
        $\beta_{1}$ & 484.53 & 1082.15 \\
        $\beta_{2}$ & $-$2201.98 & 1524.52 \\
        $\phi$      & 0.26 & 0.12 \\
        $x_{0}$     & 0.43 & 0.09 \\
        $t_{0} (= n x_{0})$ & 25.78 & 5.31 \\ \hline
        95\% CI of $t_{0}$ & \multicolumn{2}{c|}{(15.38, 36.18)} \\ \hline
    \end{tabular}
    \label{tab:ssrm_ar1_coochbehar}
\end{table}

\begin{table}[htpb]
    \centering
    \caption{Estimated change point and computation time (in milliseconds) obtained through different methods for monsoon rainfall in Coochbehar}
    \begin{tabular}{|c|c|c|}
        \hline
        \textbf{Method} & \textbf{Estimated Change Point} & \textbf{Computation Time} \\ \hline
        BR & 1991 & 44.47 \\
        SNH & 1991 & 91.61 \\
        Pettitt's & 1991 & 0.18 \\
        SeqMK & No Change Point & 84.49 \\
        SRM & 1985 & 11.50 \\
        MSR-CP (IID Error) & 1986 & 5.66 \\
        MSR-CP (AR(1) Error) & 1986 & 43.80 \\ \hline
    \end{tabular}
    \label{tab:cpt_coochbehar}
\end{table}

\section{Conclusion}    \label{sec:conclusion}
Change point analysis plays a crucial role in understanding climate data, particularly rainfall patterns. This study focuses on detecting change points in precipitation data, which is essential for assessing climate change impacts and developing effective adaptation strategies. The importance of change point estimation in climate research has been widely recognized, as it allows researchers to identify significant shifts in environmental variables over time. These shifts can have profound implications for ecosystems, agriculture and water resource management, making accurate detection methods crucial for informed decision-making.

Segmented regression has gained popularity in various fields due to its ability to model nonlinear relationships and handle data with distinct regimes. It is particularly useful for analyzing Interrupted Time Series and detecting threshold effects in environmental data. The motivation for developing this segmented regression approach stems from the limitations of traditional linear regression models in capturing changes of trends in data. But the segmented regression (\cite{fong2017a}) exhibits a sharp transition of curve at the change point of the data which is rare in real-life situations like precipitation data. Additionally, the curve is not differentiable at the change point. This restricts the use of derivative-based optimization procedures to minimize the objective function. Consequently, grid-search based binary segmentation is applied to minimize the sum of squares due to errors and estimate the change point. This grid-search based technique is computationally intensive and time consuming, particularly for large dataset.

To address these issues, a modified segmented model is discussed, which is able to capture the changing curvature in the data by incorporating a smooth, differentiable logistic sigmoid function in the segmented model. By providing a smoother transition at the change point, the modified model, namely MSR-CP, offers a more nuanced representation of rainfall trends, potentially leading to improved predictions and more accurate assessments of long-term climate patterns. This model is a particular case of smooth trend regression (STR) model discussed in various existing literature. The estimation procedure and the asymptotic properties of the estimated change point are also discussed here. The derivative-based L-BFGS-B method requires lower computation time than grid-search based binary segmentation. Also, as the model is differentiable at the change point, asymptotic normality of the estimated change point holds and consequently, an asymptotic confidence interval for the change point is obtained.

This paper extends the existing study by incorporating autoregressive errors in MSR-CP model. A joint extension of Cochrane-Orcutt transformation and Prais-Winston transformation is proposed to handle the temporal dependence in the data. The estimation procedure and the asymptotic theories are also developed. Then the method for AR(1) errors is demonstrated by simulation studies and real data analysis.

Extensive simulation studies demonstrate that the modified segmented regression model for change point show consistent estimation of change points and model parameters, with decreasing bias and standard error as sample size increases. The coverage probabilities of 95\% asymptotic confidence confidence intervals are also getting closer to 0.95 with increasing sample size. With much lower computational time, MSR-CP exhibits similar performance relative to SRM, for both true and alternative data generating scenarios. When the data are generated from MSR-CP, Buishand's Range test, Standard Normal Homogeneity test and Pettitt's test produce low biased estimates of change points but with higher standard errors. They do no provide competitive result when data are generated using alternative data generating models. Sequential Mann-Kendall test does not perform well across all scenarios. These findings highlight the reliability and versatility of modified segmented regression for change point, making it a valuable tool for researchers and policymakers working with climate data.

We analyzed real precipitation data from two districts in West Bengal, India, spanning 60 years (1961--2020). The analysis revealed change points in the mid-1980s for both districts, indicating a shift from increasing to decreasing rainfall trends. This discovery underscores the importance of change point detection in understanding regional climate patterns and their potential impacts on local communities and ecosystems. The identification of these change points can inform adaptation strategies and help stakeholders prepare for future climate-related challenges.

However, the estimated value of the change point may critically depend on the initial change point value provided in the algorithm. The process may converge to a local minimum instead of the global minimum, which is a limitation for most of the iterative algorithm. It is so advised to start with different initial values of the change point and compare the resulting change points, for practical purpose. The study can be extended further in several directions. Multiple change point can be accommodated in the MSR-CP model. Climate related variables which affect the rainfall in a region, such as temperature, wind speed and direction, humidity, particulate matter concentration in the air etc. can be included in the model to get better inference about the change point. Exact estimation algorithm can be studied further for general AR($p$) errors where $p > 1$.

In conclusion, the application of our methods to real-world precipitation data shows their practical relevance for climate research, enhancing the understanding of climate variability and improving long-term forecasting models. By detecting change points, especially in rainfall data, researchers can support water management, agricultural planning and adaptation strategies, both locally and globally. As climate change alters weather patterns, the versatility and robustness of our model, demonstrated through simulations and real-data analysis, make it a valuable tool for examining various climate variables. Our study advances the field of climate change research by providing statistical methods for detecting change points in precipitation data, combining segmented regression with a smooth continuous model. This comprehensive approach enhances the understanding of complex rainfall patterns. As climate change poses significant challenges, accurate detection of change points is crucial for informed decision-making and effective adaptation strategies.

\section*{Acknowledgment}
The authors would like to thank West Bengal Pollution Control Board for providing access to the dataset used in this article.

\section*{Funding Statement}
The authors have nothing to report.

\section*{Conflict of Interest Statement}
The authors declare no conflicts of interest.

\printbibliography

\appendix
\section{Appendix}
\subsection{Exploratory Analysis for Coochbehar Data}   \label{app:eda_coochbehar}
In this section, we conduct an exploratory analysis of the monsoon precipitation series for Coochbehar to identify underlying patterns, trends and structural features in the data, which subsequently motivates the change point analysis performed in later sections of this article. During the period 1961--2020, the mean monsoon rainfall in Coochbehar is found to be 2597.61 mm, with a standard deviation of 700.36 mm, indicating substantial inter-annual variability. The highest precipitation was recorded in 1990, amounting to 4826.06 mm, while the lowest rainfall occurred in 1994, with a total of 1339.21 mm.

\begin{figure}[htpb]
    \centering
    \includegraphics{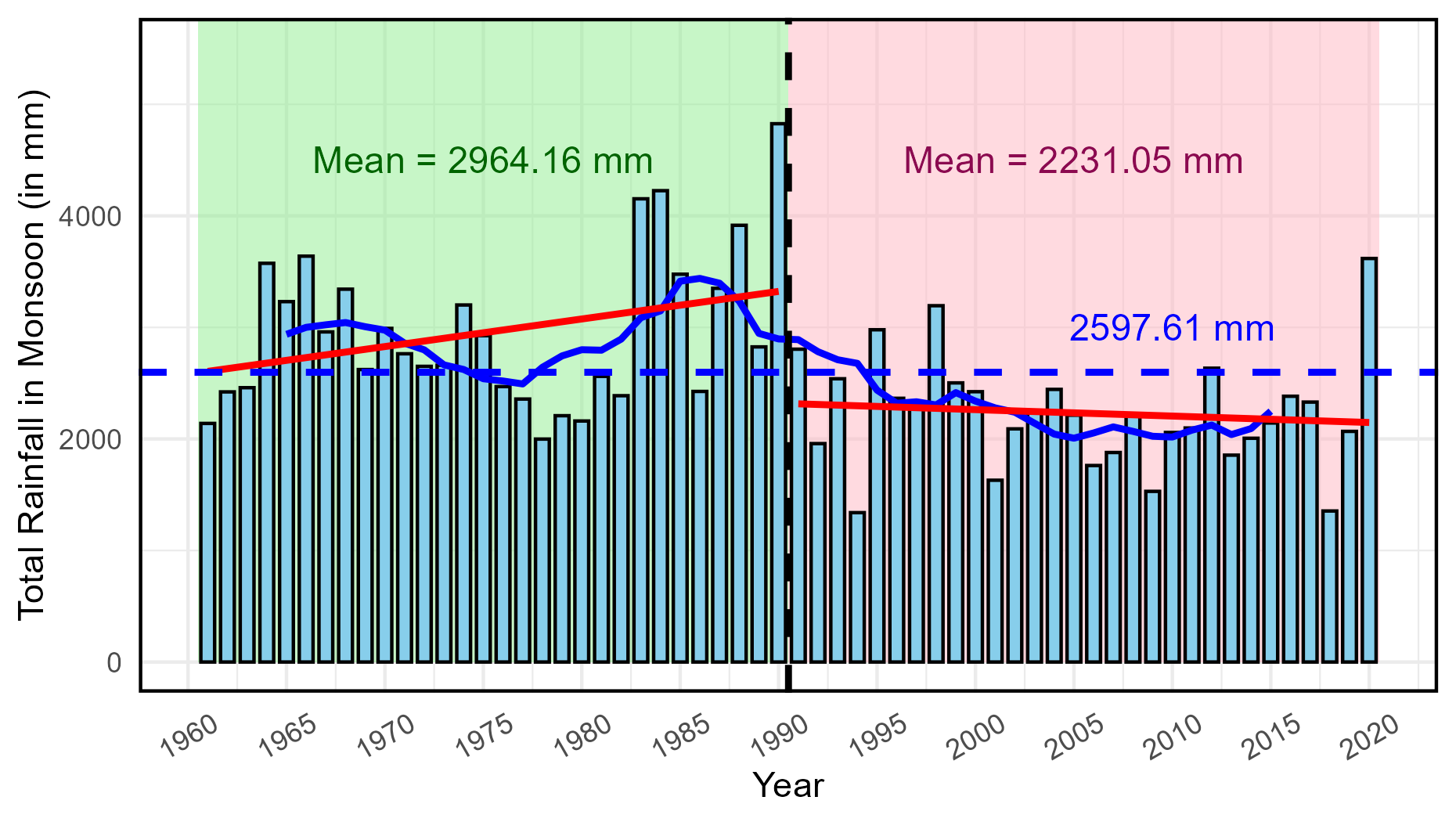}
    \caption{Yearly total precipitation (in mm) in monsoon in Coochbehar. The blue curve indicates 10-year rolling means. The red lines indicate linear trends before and after 1990. The blue dashed line is the overall mean throughout 1961--2020.}
    \label{fig:trend_coochbehar}
\end{figure}

Figure \ref{fig:trend_coochbehar} depicts the temporal evolution of monsoon rainfall across 60 years of the study period. A pronounced declining tendency in rainfall is evident after 1990, suggesting a possible structural change in the rainfall regime. Motivated by this observation, summary statistics are reported separately for the sub-periods 1961--1990 and 1991--2020 in Table \ref{tab:statistics_coochbehar}. The results reveal a substantial reduction of approximately 24.73\% in mean monsoon rainfall during 1991--2020 compared to 1961--1990. Furthermore, both the maximum and minimum rainfall levels exhibit noticeable declines in the later period, reinforcing the evidence of a downward shift in rainfall magnitude.

\begin{figure}[htpb]
    \centering
    \includegraphics{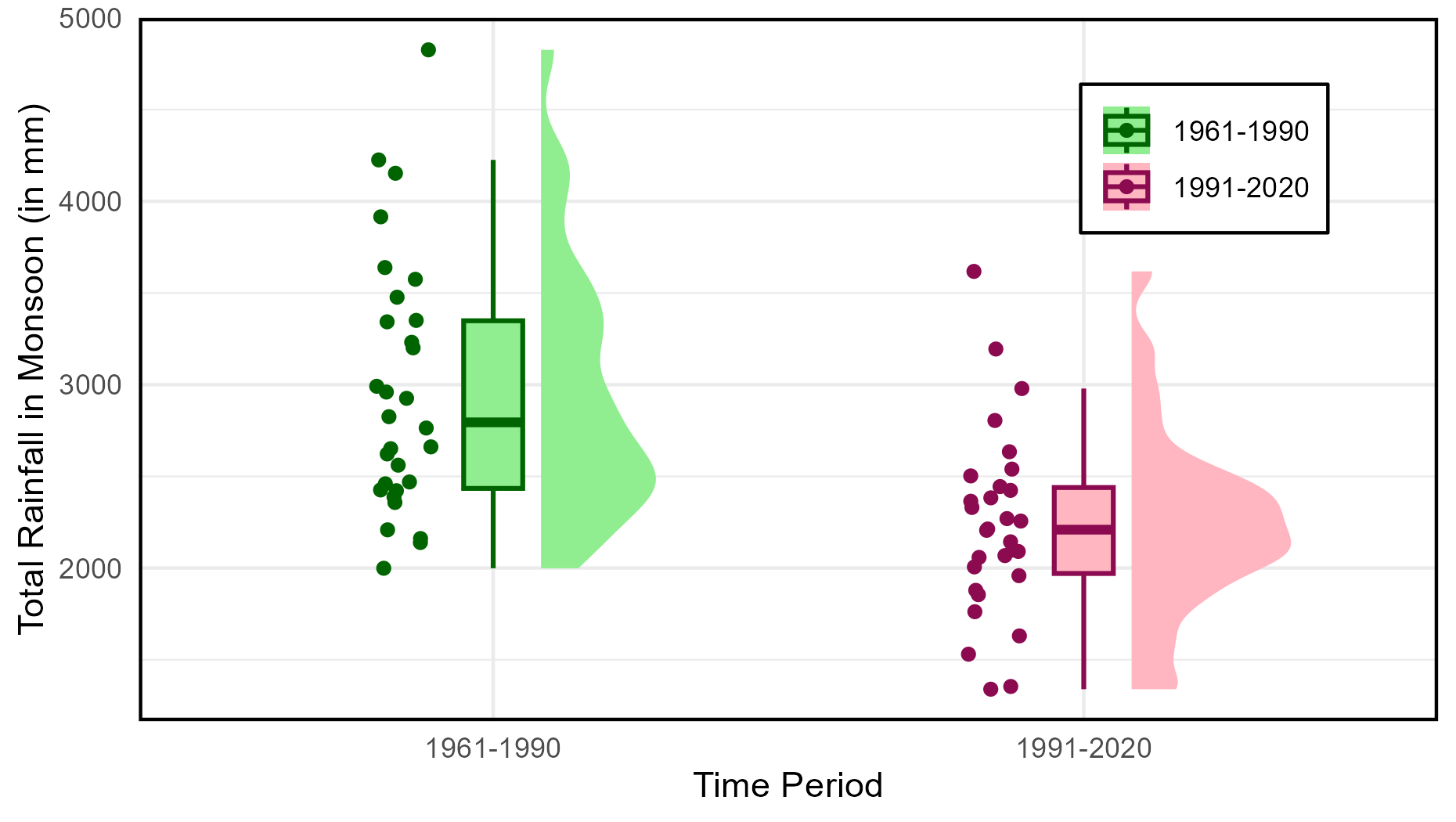}
    \caption{Distribution of annual monsoon rainfall for two sub-periods, 1961--1990 and 1991--2020, in Coochbehar district}
    \label{fig:raincloud_coochbehar}
\end{figure}

The distributional characteristics of monsoon rainfall across different time periods are further examined using raincloud plots in Figure \ref{fig:raincloud_coochbehar}. These plots clearly demonstrate a downward shift in the rainfall distribution during the most recent three decades of the six-decade study period, indicating an overall reduction in rainfall intensity and variability. This analysis indicates the existance of a change point around 1990.

\begin{table}[htpb]
\centering
\caption{Summary statistics of annual monsoon rainfall (in mm) for two sub-periods, 1961--1990 and 1991--2020, in Coochbehar district}
\begin{tabular}{|c|c|c|}
\hline
\textbf{Statistics} & \textbf{1961--1990} & \textbf{1991--2020} \\ \hline
Mean               & 2964.16 & 2231.06 \\ 
Standard Deviation & 683.89 & 494.55 \\ 
Maximum (Year)     & 4826.06 (1990) & 3617.83 (2020) \\ 
Minimum (Year)     & 1998.78 (1978) & 1339.21 (1994) \\
\hline
\end{tabular}
\label{tab:statistics_coochbehar}
\end{table}

\subsection{Proof of Theorem \ref{thm:consistency_iid}}     \label{app:consistency_iid}
\begin{align*}
    \frac{1}{n} S_{n}(\boldsymbol{\theta}) &= \frac{1}{n} \sum_{t = 1}^{n} \big[ Y_{t} - f_{t}(\boldsymbol{\theta}) \big]^{2} = \frac{1}{n} \sum_{t = 1}^{n} \big[ Y_{t} - f_{t}(\boldsymbol{\theta^{*}}) + f_{t}(\boldsymbol{\theta^{*}}) - f_{t}(\boldsymbol{\theta}) \big]^{2} \\
    &= \frac{1}{n} \sum_{t = 1}^{n} \big[ \varepsilon_{t} + f_{t}(\boldsymbol{\theta^{*}}) - f_{t}(\boldsymbol{\theta}) \big]^{2} \\
    &= \frac{1}{n} \sum_{t = 1}^{n} \Big[ \varepsilon_{t}^{2} + 2 \varepsilon_{t} \big( f_{t}(\boldsymbol{\theta^{*}}) - f_{t}(\boldsymbol{\theta}) \big) + \big( f_{t}(\boldsymbol{\theta^{*}}) - f_{t}(\boldsymbol{\theta}) \big)^{2} \Big] \\
    &= \underbrace{\frac{1}{n} \sum_{t = 1}^{n} \varepsilon_{t}^{2}}_{C_{1}} + 2 \cdot \underbrace{\frac{1}{n} \sum_{t = 1}^{n} \Big[ \varepsilon_{t} \big( f_{t}(\boldsymbol{\theta^{*}}) - f_{t}(\boldsymbol{\theta}) \big) \Big]}_{C_{2}} + \underbrace{\frac{1}{n} \sum_{t = 1}^{n} \big[ f_{t}(\boldsymbol{\theta^{*}}) - f_{t}(\boldsymbol{\theta}) \big]^{2}}_{C_{3}}
\end{align*}

By Weak Law of Large Numbers, $C_{1} \xrightarrow{p} \sigma^{2}$, as $\varepsilon_{t}$'s are i.i.d.. Similarly, $C_{2} \xrightarrow{p} 0$, as $f_{t}(\boldsymbol{\theta})$ is fixed and bounded for all $\boldsymbol{\theta} \in \boldsymbol{\Theta}$. Also, as $f_{t}(\boldsymbol{\theta})$ is uniformly continuous and bounded for $\boldsymbol{\theta} \in \boldsymbol{\Theta}$, $C_{3}$ converges uniformly to a bounded continuous function, say, $D(\boldsymbol{\theta}, \boldsymbol{\theta^{*}})$.

Now by (A2), $\beta_{2} \ne 0$. So the model (\ref{equ:model_ssrm_scaled}) is identifiable. So
\begin{equation}
    D(\boldsymbol{\theta}, \boldsymbol{\theta^{*}}) = 0 \quad \Leftrightarrow \quad \boldsymbol{\theta} = \boldsymbol{\theta^{*}}
\end{equation}

As $D(\boldsymbol{\theta}, \boldsymbol{\theta^{*}})$ cannot be negative, $D(\boldsymbol{\theta}, \boldsymbol{\theta^{*}})$ has a unique minimum at $\boldsymbol{\theta^{*}}$. So
\begin{equation*}
    \frac{1}{n} S_{n}(\boldsymbol{\theta}) \xrightarrow{p} \sigma^{2} + D(\boldsymbol{\theta}, \boldsymbol{\theta^{*}}) = Q(\boldsymbol{\theta}) \text{, say}
\end{equation*}
where $Q(\boldsymbol{\theta})$ has a unique minimum at $\boldsymbol{\theta^{*}}$. So
\begin{equation*}
    \arg \min_{\boldsymbol{\theta}} \frac{1}{n} S_{n}(\boldsymbol{\theta}) \xrightarrow{p} \arg \min_{\boldsymbol{\theta}} Q(\boldsymbol{\theta}) \quad (\text{by continuous argmin theorem}) \\
\end{equation*}
and hence
\begin{equation*}
    \boldsymbol{\hat{\theta}_{n}} \xrightarrow{p} \boldsymbol{\theta^{*}}.
\end{equation*}

\subsection{Proof of Theorem \ref{thm:asympnorm_iid}}   \label{app:asympnorm_iid}
By definition of $\boldsymbol{\hat{\theta}_{n}}$,
\begin{equation*}
    \frac{\partial S_{n}(\boldsymbol{\hat{\theta}_{n}})}{\partial \boldsymbol{\theta}} = \boldsymbol{0}.
\end{equation*}

$S_{n}(\boldsymbol{\theta})$ is continuously differentiable for infinite times. So using mean value theorem, we can write
\begin{equation*}
    \boldsymbol{0} = \frac{\partial S_{n}(\boldsymbol{\theta^{*}})}{\boldsymbol{\partial \theta}} + \frac{\partial^{2} S_{n}(\boldsymbol{\tilde{\theta}_{n}})}{\partial \boldsymbol{\theta} \partial \boldsymbol{\theta^{\top}}} (\boldsymbol{\hat{\theta}_{n}} - \boldsymbol{\theta^{*}})
\end{equation*}
where $\boldsymbol{\tilde{\theta}_{n}}$ lies between $\boldsymbol{\hat{\theta}_{n}}$ and $\boldsymbol{\theta^{*}}$. So
\begin{equation*}
    \sqrt{n} (\boldsymbol{\hat{\theta}_{n}} - \boldsymbol{\theta^{*}}) = - \underbrace{\bigg[ \frac{1}{n} \frac{\partial^{2} S_{n}(\boldsymbol{\tilde{\theta}_{n}})}{\partial \boldsymbol{\theta} \partial \boldsymbol{\theta^{\top}}} \bigg]^{-1}}_{E_{1}} \underbrace{\frac{1}{\sqrt{n}} \frac{\partial S_{n}(\boldsymbol{\theta^{*}})}{\boldsymbol{\partial \theta}}}_{E_{2}}
\end{equation*}

Now the idea is to show that $E_{1}$ converges in probability to a non-singular matrix and $E_{2}$ is asymptotically normal. We consider $E_{2}$ first.
\begin{align*}
    & \frac{\partial S_{n}(\boldsymbol{\theta})}{\boldsymbol{\partial \theta}} = -2 \sum_{t = 1}^{n} \big( Y_{t} - f_{t}(\boldsymbol{\theta}) \big) \frac{\partial f_{t}(\boldsymbol{\theta})}{\partial\boldsymbol{\theta}} \\
    \implies & \frac{1}{\sqrt{n}} \frac{\partial S_{n}(\boldsymbol{\theta^{*}})}{\boldsymbol{\partial \theta}} = - \frac{2}{\sqrt{n}} \sum_{t = 1}^{n} \big( Y_{t} - f_{t}(\boldsymbol{\theta^{*}}) \big) \frac{\partial f_{t}(\boldsymbol{\theta^{*}})}{\partial\boldsymbol{\theta}} = - \frac{2}{\sqrt{n}} \boldsymbol{F^{\top}}(\boldsymbol{\theta^{*}}) \boldsymbol{\varepsilon^{n \times 1}}
\end{align*}
where $\boldsymbol{F}(\boldsymbol{\theta^{*}}) = \big[ \frac{\partial f_{t}(\boldsymbol{\theta^{*}})}{\partial \boldsymbol{\theta}} \big]_{n \times 4}$. Now all elements of $\frac{\partial f_{t}(\boldsymbol{\theta})}{\partial \boldsymbol{\theta}}$ are uniformly continuous and bounded for all $\boldsymbol{\theta} \in \boldsymbol{\Theta^{*}}$, an open neighborhood of $\boldsymbol{\theta^{*}}$. So $\frac{1}{n} \boldsymbol{F^{\top}}(\boldsymbol{\theta^{*}}) \boldsymbol{F}(\boldsymbol{\theta^{*}})$ converges to a non-singular matrix $\Omega(\boldsymbol{\theta^{*}})$. So $E_{2}$ is a weighted average of i.i.d. $\varepsilon_{t}$. Then applying central limit theorem
\begin{equation*}
    \frac{1}{\sqrt{n}} \frac{\partial S_{n}(\boldsymbol{\theta^{*}})}{\boldsymbol{\partial \theta}} \xrightarrow{d} N_{4}(\boldsymbol{0}, 4 \sigma^{2} \Omega)
\end{equation*}
where $\Omega = \lim_{n \rightarrow \infty} \frac{1}{n}\sum_{t = 1}^{n} \frac{\partial f_{t}(\boldsymbol{\theta^{*}})}{\partial \boldsymbol{\theta}} \frac{\partial f_{t}(\boldsymbol{\theta^{*}})}{\partial \boldsymbol{\theta^{\top}}} = \lim_{n \rightarrow \infty} \frac{1}{n} \boldsymbol{F^{\top}}(\boldsymbol{\theta^{*}}) \boldsymbol{F}(\boldsymbol{\theta^{*}})$.

Now for $E_{1}$,
\begin{align*}
\frac{\partial^{2}S_{n}(\boldsymbol{\theta})}
{\partial\boldsymbol{\theta}\,\partial\boldsymbol{\theta}^{\top}}
&=
2\sum_{t = 1}^{n}
\frac{\partial f_{t}(\boldsymbol{\theta})}{\partial\boldsymbol{\theta}}
\frac{\partial f_{t}(\boldsymbol{\theta})}{\partial\boldsymbol{\theta}^{\top}}
-
2\sum_{t=1}^{n}
\bigl(Y_{t}-f_{t}(\boldsymbol{\theta})\bigr)
\frac{\partial^{2}f_{t}(\boldsymbol{\theta})}
{\partial\boldsymbol{\theta}\,\partial\boldsymbol{\theta}^{\top}}
\\
\implies
\frac{1}{n}
\frac{\partial^{2}S_{n}(\tilde{\boldsymbol{\theta}}_{n})}
{\partial\boldsymbol{\theta}\,\partial\boldsymbol{\theta}^{\top}}
&=
\underbrace{
\frac{2}{n}\sum_{t=1}^{n}
\frac{\partial f_{t}(\tilde{\boldsymbol{\theta}}_{n})}
{\partial\boldsymbol{\theta}}
\frac{\partial f_{t}(\tilde{\boldsymbol{\theta}}_{n})}
{\partial\boldsymbol{\theta}^{\top}}
}_{G_{1}}
\\
&\quad
-
\underbrace{
\frac{2}{n}\sum_{t=1}^{n}
\bigl(f_{t}(\boldsymbol{\theta}^{*})
-f_{t}(\tilde{\boldsymbol{\theta}}_{n})\bigr)
\frac{\partial^{2}f_{t}(\tilde{\boldsymbol{\theta}}_{n})}
{\partial\boldsymbol{\theta}\,\partial\boldsymbol{\theta}^{\top}}
}_{G_{2}}
\\
&\quad
-
\underbrace{
\frac{2}{n}\sum_{t=1}^{n}
\varepsilon_{t}
\frac{\partial^{2}f_{t}(\tilde{\boldsymbol{\theta}}_{n})}
{\partial\boldsymbol{\theta}\,\partial\boldsymbol{\theta}^{\top}}
}_{G_{3}}.
\end{align*}

Now $\boldsymbol{\hat{\theta}_{n}}$ is consistent for $\boldsymbol{\theta^{*}}$. So $\boldsymbol{\tilde{\theta}_{n}} \xrightarrow{p} \boldsymbol{\theta^{*}}$. So $\boldsymbol{\tilde{\theta}_{n}} \in \boldsymbol{\Theta^{*}}$. Also all elements of $\frac{1}{n} \sum_{t = 1}^{n} \frac{\partial f_{t}(\boldsymbol{\theta})}{\partial \boldsymbol{\theta}}$ and $\frac{1}{n} \sum_{t = 1}^{n} \frac{\partial^{2} f_{t}(\boldsymbol{\theta})}{\partial \boldsymbol{\theta} \partial \boldsymbol{\theta^{\top}}}$ converges uniformly for $\boldsymbol{\theta} \in \boldsymbol{\Theta^{*}}$. So by Weak Law of Large Numbers $G_{2} \xrightarrow{p} 0$ and $G_{3} \xrightarrow{p} 0$. Also $G_{1} \xrightarrow{p} 2 \Omega$. So $E_{1} \xrightarrow{p} \frac{1}{2} \Omega^{-1}$, which is a non-singular matrix. Hence
\begin{equation*}
    \sqrt{n}(\boldsymbol{\hat{\theta}_{n}} - \boldsymbol{\theta^{*}}) \xrightarrow{d} N_{4}(\boldsymbol{0}, \sigma^{2} \Omega^{-1}).
\end{equation*}

\end{document}